\documentclass{article}%
\usepackage{amsmath}
\usepackage{amsfonts}
\usepackage{amssymb}
\usepackage{graphicx}
\usepackage{subcaption}
\usepackage{multirow}
\usepackage{color}
\usepackage{longtable}
\usepackage[dvipsnames]{xcolor}
\usepackage[bookmarks=true,colorlinks=true]{hyperref}
\hypersetup{linkcolor=blue}
\hypersetup{citecolor=blue}
\newcommand{\bpartial}{\mathop{\partial\kern -4pt\raisebox{.8pt}{$|$}}}
\newcommand{\bra}{\mathopen{[\kern-1.6pt[}}
\newcommand{\ket}{\mathclose{]\kern-1.5pt]}}
\newcommand{\bbra}{\mathopen{[\kern-2.2pt[\kern-2.3pt[}}
\newcommand{\bket}{\mathclose{]\kern-2.1pt]\kern-2.3pt]}}

\makeindex
\begin{document}
	\title {\large{ \bf 
			New five-dimensional Bianchi type magnetically charged hairy topological black hole solutions in string theory}}
	
	\vspace{3mm}
		\author {  \small{ \bf  F. Naderi}\hspace{-1mm}{ \footnote{Corresponding author,
					 e-mail:
			 f.naderi@azaruniv.ac.ir}} ,{ \small	} \small{ \bf  A. Rezaei-Aghdam}\hspace{-1mm}{
		\footnote{
			 e-mail:	rezaei-a@azaruniv.ac.ir}} \\
		 		{\small{\em
			Department of Physics, Faculty of Basic Sciences, Azarbaijan Shahid Madani University}}\\
	{\small{\em   53714-161, Tabriz, Iran  }}}

\maketitle

\begin{abstract}
We construct black hole solutions to the leading order of string effective action in five dimensions with the source given by dilaton and magnetically charged antisymmetric gauge $B$-field. Presence of the considered $B$-field leads to the unusual asymptotic behavior of the solutions which are neither asymptotically flat nor asymptotically (A)dS. We consider the three-dimensional space part to correspond to the Bianchi classes and so the horizons of these topological black hole solutions are modeled by seven homogeneous Thurston’s geometries of $E^3$, $S^3$, $H^3$, $ H^2 \times E^1$, $\widetilde{{SL_2R}}$,   nilgeometry, and solvegeometry.  Calculating the quasi-local mass, temperature, entropy, dilaton charge, and magnetic potential, we show that the first law of black hole thermodynamics is satisfied by these quantities and the dilaton charge is not  independent of mass and magnetic charge. Furthermore, for Bianchi type $V$, the $T$-dual black hole solution is obtained which carries no charge associated with  $B$-field and the entropy turns to be invariant under the $T$-duality.

\end{abstract}

	\section{Introduction}
	The low energy string effective action  contains a rich variety of black hole solutions, which being characterized by  charges of  dilaton, Yang-Mills fields and the antisymmetric tensor gauge field ($B$-field), possess qualitatively different properties from those of general relativity
	\cite{GIBBONS1988741,garfinkle1991charged,shapere1991dual,witten1991string}. In this context, the rotating and charged dilaton black hole solutions in string
	theories have been obtained, for instance, with constant dilaton for the dilaton-graviton system in \cite{thorne1986black} and with including a  coupling of dilaton to the Maxwell field within a generalized class of theories that arise in low energy string
	effective action in \cite{horne1992rotating}.
	Furthermore, taking into account a non-trivial  $B$-field, or more generally the $p$-form fields, the charged and rotating string black hole solutions have been found for example in  \cite{horne1992rotating,BURGESS199575,sen1992rotating,DABHOLKAR199685,HorneHorowitz,Emparan2004,Bardoux2012}. In this category, the well known  Kerr-Sen rotating  charged black hole solution  in  low energy effective field theory of heterotic string theory  
	is the first 
	constructed solution   by transforming the Kerr solution
	using the twisting method \cite{sen1992rotating}.

	The equations of motion of  low energy string frame effective action  are  equivalent to the one-loop $\beta$-function equations which are the conformal invariance condition of the corresponding  $\sigma$-model. 
	Being differential equations, like Einstein field equations, they  characterize the local properties of spacetime and not the global structure, namely the topology of spacetime. 
	The restriction on the topology of black holes was first noticed in  Hawking's theorem where the event horizon cross section of a stationary asymptotically flat $4$-dimensional black hole, which satisfies the dominant energy condition, was determined to be $S^2$ \cite{Hawking1975,hawking1973large}. A generalization of this result to the higher dimensional black holes was then provided in \cite{Galloway2006}, where the event horizon of a stationary black hole in arbitrary dimension is required to admit a positive scalar curvature metric. The Emparan and Reall black ring solution with  $S^2 \times S^1$  horizon topology is a relevant example in this category \cite{Emparan2004}. 
	Nevertheless, the black hole solutions with non-trivial horizon topology, namely the topological black holes, are allowed to be considered if some of the assumptions of these theorems are relaxed \cite{Aminneborg,Smith,Brill,Yazadjiev}. For example, in the asymptotically anti-de Sitter (AdS) spacetimes, where the dominant energy condition and asymptotic flatness are both violated, the black hole solutions with  compact Riemann surfaces  of any genus
	$g $ as horizon  have been studied in \cite{LEMOS199546,Mann,Vanzo,Cai,Topologicalbh1}. Also, the rotating generalization of these solutions in AdS$_4$ with $g>0$ has been presented in \cite{Klemm,Klemm2}, whose horizons  are not compactificable  and their described  spacetimes are  rotating black branes \cite{Klemm3}.

	Usually, the horizon geometries of topological black hole solutions are considered to be spherical, hyperbolic or flat. 
	For the five-dimensional black hole solutions the situation can be more extensive. The  geometries of three-dimensional homogeneous manifolds has been classified by Thurston into eight types \cite{thurston1983three}, where in addition to  three isotropic constant curvature
	cases of spherical $S^3$, Hyperbolic $H^3$, and  Euclidean $E^3$, there are five anisotropic alternative  three-dimensional geometries of  $S^2 \times R$, $H^2 \times
	R$,   $\widetilde{{SL_2R}}$, solvegeometry, and nilgeometry. 
	Admitting homogeneous
	metrics, these model geometries have close  correspondence with the   homogeneous geometries of
	Bianchi types and Kantowski-Sachs  \cite{Closed,closedBianchi}.
	The Bianchi types are  defined based on the simply-transitive
	three-dimensional Lie groups classification. 
	Since the isometries of the Riemannian manifold form a Lie group, the
	Bianchi classification is widely used for  studying the spatially
	homogeneous spacetimes in relativity and cosmology   \cite{Ellis1969,batakis2,PhysRevD.57.5108,NADERI2017,Naderi2}. The spacetimes in  these  models    are  assumed  to  possess a symmetry, namely the spatial homogeneity  \cite{Cosmictopology}.
	Families of five-dimensional black hole solutions of gravity theories where the horizons are modeled by some of the Thurston  $3$-geometries have been presented in \cite{Cadeau,Liu2012,PhysRevD.91.084054,PhysRevD.97.024020,HERVIK20081253}. Also, extremal black branes whose geometries are classified by nine Bianchi types have been studied in \cite{Iizuka2012}.
	
	Recently, the black hole solutions in string theory have  attracted
	an increasing interest \cite{PhysRevD80104032,PhysRevD83044002,Moura2013,PhysRevD94104067,Maeda2010,Quintin,EGHBALI2017791}.
	Nevertheless, the effect of the magnetically charged $B$-field has not been discussed so
	much in previous works  \cite{BURGESS199575,horne1992rotating,sen1992rotating,DABHOLKAR199685,HorneHorowitz,Emparan2004}.  
	In this work, we are interested in five-dimensional black hole solutions in string theory including a  dilaton field and a $3$-form field strength tensor of $B$-field associated with a magnetic charge. 
	The inclusion of $B$-field in this class will affect the asymptotic properties of the black hole spacetime where the asymptotic flatness condition can be violated, which brings up the possibility of considering the unusual geometries. 
	The three-dimensional horizons will be assumed to be  homogeneous spaces corresponding to  Bianchi types with closed spatial sections.
	
	Generally, presence of a non-trivial dilaton, which introduces a dilaton charge, leads to dependence of mass on the asymptotic value of dilaton at infinity, $\phi_{\infty}$.  Dilaton charge has a critical role in the physics of black holes.
	The no-hair theorems, for example in Brans-Dieke  theory with minimally coupled dilaton to gravity \cite{Hawking1972}, indicate that the dilaton is not allowed to carry a charge. However, in some cases, the scalar field can have a non-trivial profile  introducing a dilaton charge which is not an extra independent characteristic of the black hole. This is usually referred as the scalar hair of second kind  \cite{Sotiriou}.  For this kind of solutions, a cosmological scenario has been provided in \cite{Gibbons}, where $\phi_{\infty}$ does vary and so the scalar charge appears in the first law of thermodynamics.
	The problem with this modified first law of thermodynamics is that the dilaton charge, not being protected by a gauge symmetry,  is not a conserved charge \cite{Astefanesei1,Astefanesei2008}. This charge is  non-localized, exists entirely outside the horizon and usually corresponds to a secondary hair \cite{Gibbons,Astefanesei1}. 
Also, in the AdS spacetimes,  scalar charges are not compatible with AdS/CFT  since  no such  charge exist in dual CFT.  In fact, it is proved in complete generality that  for asymptotically AdS black holes the first law is satisfied without charge changes \cite{Papadimitriou2005}. Noting that, even though  the  energy does depend on  $\phi_{\infty}$ in general, see for instance	the   holographic  stress-energy tensor expressions (from which  the energy can be computed) in \cite{Bianchi2001,BIANCHI2002159}. 
	An intriguing physical discussion about the variation of $\phi_{\infty}$ is presented in \cite{HAJIAN2017228}, where using the solution phase space method and integrability of charge variations, it is shown that   for black hole solutions of Einstein-Maxwell-(Axion)-Dilaton theories the $\phi_{\infty}$ is a redundant (non-physical) parameter, whose  variation is not included in the first law of thermodynamics. Then consolidated by the idea that some redefinitions based on shift symmetry can eliminate the dependence of the mass and charges on $\phi_{\infty}$ \cite{Astefanesei1,Astefanesei2008}, the applicability of the result in any other theory is asserted, declaring that the $\phi_{\infty}$ is generally a redundant parameter.  
	Alternatively, inspired by the  counterterm subtraction method which has been developed  to define the conserved charges of AdS spacetimes  in   \cite{Henningson1998,Balasubramanian1999,SKENDERIS2001,deHaro2001}, it is shown in \cite{ASTEFANESEI201847} that a well defined variational principal can be obtained by including a boundary term (counterterm) for dilaton field in  Brown-York formalism which leads to a total energy with the contribution of  scalar field where the first law of thermodynamics for this  total energy does not contain the variation term of $\phi_{\infty}$. 
	Motivated by  this, we will consider an appropriate boundary term for the dilaton field and investigate its consequence on the first law of black hole thermodynamics.

	The paper is organized as follows: In section \ref{sec2}, a review on low energy $D$-dimensional string effective action and its equations of motion is presented. 
	In section \ref{sec3}, the string equations of motion are solved to obtain new five-dimensional  black hole solutions   whose horizons are homogeneous spaces corresponding to the Bianchi types. Then, the thermodynamic behavior  and the extremal condition of the solutions are investigated in section \ref{sec4}.  Furthermore, for Bianchi type $V$ case the $T$-dual solutions and their thermodynamic properties are investigated in section \ref{sect}. Finally, some concluding remarks are presented in section \ref{conclussion}.
	
	\section{  Low energy string effective action and one-loop \texorpdfstring{$\beta$}{Lg}-functions }\label{sec2}
	For a $\sigma$-model, the one-loop $\beta$-function equations  of the  background fields of metric $g_{\mu\nu}$, dilaton $\phi$ and antisymmetric tensor gauge $B$-field 
	are given by \cite{FRADKIN19851,callan1985strings}
	\begin{equation}\label{betaGR}
	\begin{aligned}
	{R}_{{\mu}{\nu}}-\frac{1}{4}{H}^{2}_{{\mu}{\nu}}-\nabla_{{\mu}}\nabla_{{\nu}}{\phi}=0,
	\end{aligned}
	\end{equation}
	\begin{equation}\label{betab}
	\begin{aligned}
	\nabla^{\mu}\left({\rm e}^{\phi}H_{\mu\nu\rho}\right)=0,
	\end{aligned}
	\end{equation}
	\begin{eqnarray}\label{betafR}
	{R}-\frac{1}{12}{H}^{2}+2\nabla_{{\mu}}\nabla^{{\mu}}{\phi}+(\partial_{{\mu}}{\phi})^{2}-\Lambda=0,
	\end{eqnarray}
	where  $H_{\mu\nu}^2=H_{\mu\rho\sigma}H_{\nu}^{\rho\sigma}$	and $H$ is  the field strength of $B$-field defined by $H_{\mu\nu\rho}=3\partial_{[\mu}B_{\nu\rho]}$.
	These equations can be also obtained 	
	by variation of the following string frame effective action with respect to  metric, $B$-field and  dilaton 
	\begin{eqnarray}\label{action}
	S=\frac{-1}{2 \lambda_s^{D-2}}\int d^D x\sqrt{g}{\rm e}^{\phi}(R-\frac{1}{12}H^2+(\nabla\phi)^2+\Lambda),
	\end{eqnarray}	
	where $\lambda_s$ is the string length and the $\Lambda$  is related to the central charge deficit of theory which is  given  in non-critical $D$-dimensional bosonic theory  as $	\Lambda=\frac{2\,(26-D)}{3\alpha'}$ \cite{tseytlin1992elements}.
	We
	will set $\Lambda=0$ in our analysis. This may be a good approximation  if  the
	curvature and/or kinetic energy are large compared to the $\Lambda$, i.e. $\Lambda\ll R, \nabla\phi, H^2$ \cite{Copeland}.
	Alternatively,  the   Einstein frame can be introduced whose  metric $\tilde{g}_{\mu\nu}$  is related to the string frame metric ${g}_{\mu\nu}$ in $D$-dimensional spacetime by
	\begin{equation}\label{con}
	\tilde{	g}_{\mu\nu}={\rm e}^{\frac{2}{D-2}\phi}g_{\mu\nu}.
	\end{equation}	 
	Performing this conformal transformation,  
	the  Einstein frame effective action   for bosonic string can be obtained as \cite{FRADKIN19851,callan1985strings}
	\begin{eqnarray}\label{GBaction}
	\begin{split}
	S=-\frac{1}{2\kappa_D^2}\int d^D x\sqrt{\tilde{	g}}\bigg(&\tilde{R}
	-\frac{(\tilde{\nabla}\phi)^2}{D-2}-\frac{1}{12}\,{\rm e}^{\frac{4\phi}{D-2}}H^2\bigg),
	\end{split}
	\end{eqnarray}
	in which  $\tilde{\nabla}$ indicates the covariant derivative with respect to  $\tilde{g}$, and
	$\kappa_D^2=8\pi G_D=\lambda_s^{D-2}{\rm e}^{-\phi}=\lambda_p^{D-2}$, where $\lambda_p$ is the Planck length and 
	$G_D$ is the $D$-dimensional gravitational Newton constant.
	Principally,  understanding the gravitational phenomena is more convenient in Einstein frame in which the dilaton coupling to Ricci scalar in \eqref{action} has been eliminated. 
	In this frame, the one-loop $\beta$-functions \eqref{betaGR}-\eqref{betafR} can be rewritten in the following form of Einstein field equations that can be also obtained by  the variation of  the  action \eqref{GBaction} with respect to $\tilde{    g}_{\mu\nu}$  \cite{callan1985strings}
	\begin{eqnarray}\label{22}
	\begin{split}
	\tilde{R}_{\mu\nu}-\frac{1}{2}	\tilde{R}	\tilde{g}_{\mu\nu}=\kappa_D^2T_{\mu\nu}^{\mathrm{(eff)}},
	\end{split}
	\end{eqnarray}
	where the effective  energy-momentum tensor is defined as follows 
	\begin{eqnarray}\label{Teff}
	\begin{split}
	T_{\mu\nu}^{\mathrm{(eff)}}	=T_{\mu\nu}^{(\phi)}+T_{\mu\nu}^{(B)},
	\end{split}
	\end{eqnarray}
	and
	\begin{eqnarray}
	\kappa_D^2T_{\mu\nu}^{(\phi)}=\frac{1}{D-2}(\tilde{\nabla}_{\mu}\phi\tilde{\nabla}_{\nu}\phi-\frac{1}{2}\,\tilde{g}_{\mu\nu}(\tilde{\nabla}\phi)^2),\label{Tfi}\\
	\kappa_D^2T^{(B)}_{\mu\nu}=\frac{{\rm e}^{\frac{4\phi}{D-2}}}{4}(H_{\mu\kappa\lambda}H_{\nu}^{\kappa\lambda}-\frac{1}{6}H^2\tilde{g}_{\mu\nu})\label{TB},
	\end{eqnarray}
	Also, the equations of motion of dilaton and $B$-field  are given by \cite{callan1985strings}
	\begin{eqnarray}\label{}
	\tilde{\nabla}^2\phi-\frac{1}{12}{\rm e}^{\frac{4\phi}{D-2}}H^2=0,\label{eq1}\\
	\tilde{\nabla}_{\mu}\left({\rm e}^{\frac{4\phi}{D-2}}H^{\mu}_{~\nu\rho}\right)=0.\label{eq2}
	\end{eqnarray}

	\section{Black hole solutions of  equations of motion of low energy string theory  in five dimensions} \label{sec3}
	In this paper we focus on $D=5$ case, looking for black hole
	solutions of  low energy string theory effective equations of motion  on five-dimensional spacetimes where  the $r$ and $t$ constant hypersurfaces are given by homogeneous space corresponding to a Bianchi types. In this regard, we start with the string frame metric ansatz 
	\begin{eqnarray}\label{metric}
	ds^2=g_{\mu\nu}\,dx^{\mu}\,dx^{\nu}=-F\,dt^{2}+ \frac{1}{F\,U^2}\,dr^{2}+g_{i}^2\sigma^{i}\sigma^{i},
	\end{eqnarray}
	in which the metric components are function of  radial coordinate, $ F=F(r)$, $U=U(r)$, and $g_{i}=g_{i}(r)$, where $i=1, 2, 3$ and $\sigma^i$ are left invariant basis $1$-forms. The Bianchi type classification and their left invariant $1$-forms are presented in Appendix \ref{app1}.    Respecting the homogeneity of spacetime, the dilaton field is taken to be a  function of $r$ only. 
	The contribution of field strength tensor $H$  for Bianchi classes  with diagonal metrics can be classified based on the
	orientation of  its 
	Hodge dual with respect to the three-dimensional
	hypersurface of homogeneity  sections. Here,
	the field strength tensor of $B$-field, satisfying $dH=0$, is chosen to be in the following class \cite{batakis2}
	\begin{eqnarray}\label{H}
	H=\frac{1}{3!}\,b\,\sigma^{1}\wedge\sigma^{2}\wedge\sigma^3,
	\end{eqnarray}
	where $b$ is a constant. 
	Then, using  \eqref{rnon}-\eqref{gnon},
	the $(t,t)$, $(r,r)$ and $(i,i)$ components of  $\beta$-function of metric \eqref{betaGR} and  $\beta$-function of dilaton \eqref{betafR}  
	reduce to following coupled set of differential equations  
	\begin{eqnarray}\label{tt}
	-\frac{1}{2}\frac{F''}{F}-\frac{1}{2}\ln F'(\sum \ln g_i'+\phi'+\ln U')=0,
	\end{eqnarray}
	\begin{eqnarray}\label{rr}
	\begin{aligned}
	-\phi''&-\sum\frac{g_i''}{g_i}-\frac{1}{2}\frac{F''}{F}-\frac{1}{2}\ln U'\ln F'-\frac{1}{2}(2\ln U'+\ln F')(\sum \ln g_i'+\phi')=0,
	\end{aligned}
	\end{eqnarray}
	\begin{eqnarray}\label{ii}
	\begin{aligned}
	-\frac{g_i''}{g_i}&-\ln g_i'(\sum_{j\neq i}\ln g_j'+\phi'+\ln U'+\ln F')-\frac{1}{2}b^2(g_1g_2g_3U)^{-2}F^{-1}-Y_i=0,
	\end{aligned}
	\end{eqnarray}
	\begin{eqnarray}\label{fi}
	\begin{aligned}
	2{\phi}''+\sum(2\,\frac{g_i''}{g_i}+\frac{F_i''}{F_i}&+2 \ln g_i'(\phi'+\ln F'+\ln U')+Y_i)+2\phi'(\ln U'+\ln F'+\frac{1}{2}\phi')+\ln U'\ln F'\\
	&+2\sum_{i<j}\ln g_i' \ln g_j'+\frac{1}{2}b^2(g_1g_2g_3U)^{-2}F^{-1}=0,
	\end{aligned}
	\end{eqnarray}
	where the prime stands for   derivative with respect to $r$ and the $Y_i$ terms are Bianchi type dependent terms which will
	be given in the following subsections.
	Furthermore, the equation \eqref{betaGR} imposes the the following constraint equations in Bianchi types of class $B$ which possess  non-diagonal components of Ricci tensor
	\begin{eqnarray}\label{classBc}
	R_{\mu\nu}=0,\quad (\mu\neq \nu).
	\end{eqnarray}
	Now, adding  $\eqref{tt}$, \eqref{rr}, and  summed over $i$ of \eqref{ii}   to  \eqref{fi} leads to the following equation
	\begin{eqnarray}\label{33}
	{\phi}''+{\phi}'(\sum \ln g_i'+\ln (FU{\rm e}^{\phi})')-b^2(g_1g_2g_3U)^{-2}F^{-1}=0.
	\end{eqnarray}
	Also, adding twice of \eqref{rr} to \eqref{fi}   yields the initial value equation as follows
	\begin{eqnarray}\label{ini}
	\begin{aligned}
	\phi'(\phi'+\ln F')&+\sum \ln g_i'(2 \phi'+\ln F')+2\,\sum_{j\neq i}\ln g_i'\ln g_j'+\sum Y_i+\frac{1}{2}b^2(g_1g_2g_3U)^{-2}F^{-1}=0.
	\end{aligned}
	\end{eqnarray}
	
	In order to have black hole solutions with regular horizon $r_H$,  we impose  the following boundary conditions  
		\begin{eqnarray}
		F_H=0,\quad F_H'\neq 0,
		\end{eqnarray}
		with finite $g_{iH}$, $ \phi_H$ and $U_H$. Here and in what follows, the subscript $H$ indicates the quantities evaluated at  the horizon. Also, from \eqref{33} we get
		\begin{eqnarray}\label{bof}
		\phi'_{H}F'_{H}=\frac{b^2}{(g_{1H}g_{2H}g_{3H}U_H)^2}.
		\end{eqnarray}
		In the asymptotic region $r\rightarrow\infty$, we are interested in non-logarithmic branch with $F(r)=c+{\cal{O}}(r^{-1})$ and
		\begin{eqnarray}
		\phi\sim \phi_{\infty}+\frac{{\cal{D}}}{r}+\frac{{\cal{C}}}{r^2},\label{fiex}
		\end{eqnarray}
		where  $c$, ${\cal{D}}$ and ${\cal{C}}$  are finite constants. The relevant asymptotic behavior for metric functions are
		\begin{eqnarray}
		U\sim r^2\left(u_{\infty}+\frac{u_1}{r}\right),\label{ue}\\
		g_i\sim g_{i\infty}+\frac{g_{i1}}{r}\label{ge},
		\end{eqnarray}
		where  $u_{\infty}$, $u_{1}$, $g_{i\infty}$ and $g_{i1}$ are finite constants.

	To solve the set of equations of \eqref{tt}, \eqref{rr}, and \eqref{33} subject to the  initial value equation \eqref{ini}, we choose  
	\begin{eqnarray}\label{f}
	U(r)=r^2(g_1g_2g_3)^{-1}{\rm e}^{-\phi}.
	\end{eqnarray} 
	In this case, the equations \eqref{tt},  and \eqref{33} read
	\begin{eqnarray}\label{h}
	F''+2\,r^{-1}F'=0,
	\end{eqnarray}
	\begin{eqnarray}\label{ff}
	\phi''+\phi'\ln(r^2\,F)'-\frac{b^2}{r^4F}{\rm e}^{2\,\phi}=0.
	\end{eqnarray}
	Also, using \eqref{fi} in \eqref{ii} we get
	\begin{eqnarray}\label{newii}
	\ln(g_i^2{\rm e}^\phi)''+\ln(g_i^2{\rm e}^\phi)'\ln(r^2\,F)'+2\,Y_i=0.
	\end{eqnarray}
	The solutions of \eqref{h} and \eqref{ff} give the $F(r)$  and $\phi(r)$, independent of the type of three-dimensional Bianchi part, as follows
	\begin{eqnarray}\label{45}
	F=c_2+\frac{c_1}{r},
	\end{eqnarray}
	\begin{eqnarray}\label{fioa}
	\phi=-\ln\left({c_3^2F^{\frac{1}{2}-n}-b^2F^{\frac{1}{2}+n}}\right)+\ln(2nc_1c_3),
	\end{eqnarray}
	where the $c_1$, $c_2$, $ c_3$, and $n$ are real integrating constants. 
	
	For dilaton $\phi$ \eqref{fioa} and $\phi'$, which is  given by 
		\begin{eqnarray}
		\phi'=-\frac{c_1}{r^2}\frac{\left(n+\frac{1}{2}\right)b^2F^{2n}+c_3\left(n-\frac{1}{2}\right)}{F\,\left(c_3^2-b^2F^{2n}\right)},
		\end{eqnarray}
		to be regular at the  horizon and consistent with the boundary condition \eqref{bof},  the restriction of 
		\begin{eqnarray}\label{n}
		n=\frac{1}{2},
		\end{eqnarray}
		is required.
	
	To have black hole interpretation in the solutions, the $c_1$ constant is required to be negative with positive $c_2$. Also, to have well defined dilaton field given by \eqref{fioa} in the black hole solutions we  assume that 
	\begin{eqnarray}\label{condition1}
	{c_3^2-b^2}>0,\quad c_3<0,\quad c_1<0. 
	\end{eqnarray}
	In some cases of topological  black hole solutions whose horizons are  constant curvature  spaces, the requirement of solution to be asymptotically AdS spacetime, i.e. Einstein space with negative
	cosmological constant, relates the integrating constant of type $c_2$  to the curvature constant of horizon \cite{Topologicalbh1}. 
	Here, considering \eqref{con}, \eqref{22}, \eqref{metric} and \eqref{H} the components of Ricci tensor in Einstein frame are given by
	\begin{eqnarray}\label{r} 
	2\tilde{	R}_{i}^{i}=-\tilde{	R}_{t}^t=\frac{b^2\rm e^{\frac{4}{3}\phi}}{3\tilde{\gamma}},\,\tilde{	R}_{r}^{r}=\frac{-b^2{\rm e^{-\frac{4}{3}\phi}}+r^4\phi'^2F{{\rm e}^{{-\frac{2}{3}\phi}}}}{3\tilde{\gamma}},\,\,
	\end{eqnarray}
	where $\tilde{\gamma}$ is the determinant of the Einstein frame  metric of three-dimensional homogeneous space in \eqref{metric}, defined by
	\begin{eqnarray}\label{sigma}
	\tilde{\gamma}=\left(\tilde{g}_1 \tilde{g}_2 \tilde{g}_3 \right)^2=\left(g_1 g_2 g_3 \right)^2{\rm e}^{2\phi}.
	\end{eqnarray}
	Noting the \eqref{45}, \eqref{fioa} and \eqref{r} it can be checked that our solutions do not admit asymptotically Einstein spaces, with $\tilde{	R}_{\mu\nu}=k\,\tilde{	g}_{\mu\nu}$. 
	Hence,  having no explicit condition to fix the $c_2$ parameter,  we set $c_2=1$ for simplicity. The asymptotic value of dilaton \eqref{fioa} is then given by
	\begin{eqnarray}\label{fiinfini} \phi_{\infty}=\ln(\frac{c_1c_3}{c_3^2-b^2}).
	\end{eqnarray}
	
	So far, we have not considered any special kind of geometry for three-dimensional homogeneous space. The effective energy-momentum tensor \eqref{Teff} satisfies  all of the energy conditions but the Ricci scalar in Einstein frame
	\begin{eqnarray}\label{ricci} 
	\tilde{	R}=\frac{-b^2{\rm e^{-\frac{4}{3}\phi}}+2r^4\phi'^2F{{\rm e}^{{-\frac{2}{3}\phi}}}}{6\tilde{\gamma}},
	\end{eqnarray}
	has the  following  asymptotic form   in $r\rightarrow \infty$ limit
	\begin{eqnarray}\label{trace}
	\tilde{	R}=\frac{3\,b^2{\rm e}^{\frac{4}{3}\phi_{\infty}}}{2c_3^2\tilde{\gamma}_{\infty}}\left(2b^2-c_3^2\right),
	\end{eqnarray}
	where $\tilde{\gamma}_{\infty}=\lim_{r\rightarrow\infty} \tilde{\gamma}$. 
	Evidently, the presence of the field strength tensor  of type \eqref{H} has affected the asymptotic behavior of these solutions. In such a way that,
	as long as $c_3\neq-\sqrt{2}b$  and the  $\tilde{\gamma}_{\infty}$  is finite, the solutions are not either asymptotically flat or (considering the \eqref{r})  asymptotically (A)dS.\footnote{As we will see in subsections \ref{i}-\ref{ix}, all of the Bianchi type solutions have finite $\tilde{\gamma}_{\infty}$. }  Therefore, having violated asymptotically flatness condition, the considered spacetimes are not fundamentally forbidden to have negatively curved   horizon geometries. 
	In other words, besides the flat Bianchi type $I$ and positively curved Bianchi type $IX$, we can consider the other Bianchi types with negative three-dimensional curvatures.

	Determining the Bianchi type dependent terms $Y_i$ in \eqref{newii} using  relations of \eqref{rnon}-\eqref{gnon}, we are going to  find the solutions of \eqref{newii} in all Bianchi types in the following subsections to establish the form of  metric of the string and Einstein frames.

	\subsection{Bianchi type $I$}\label{typeisol}

	This Bianchi type has $Y_i=0$  and  solutions of \eqref{newii}   give the components of homogeneous part of string frame metric  \eqref{metric} as 
	\begin{eqnarray}\label{Igi}
	g_i ^2={\rm e}^{-\phi}F^{2\,p_i},
	\end{eqnarray}
	where $p_i$ are integrating constants.  Substituting them  into the initial value equation \eqref{ini} gives the following condition on  constants
	\begin{eqnarray}\label{iniI}
	\begin{aligned}
	-2\sum p_i-4\sum_{i<j}p_i \,p_j=0.
	\end{aligned}
	\end{eqnarray}
	Also, 
	the conformal transformation \eqref{con} gives the Einstein frame metric by
	\begin{eqnarray}\label{Isol}
	\begin{aligned}
	ds^2=-{ W  }^{-\frac{2}{3}}{ F} \,dt^2+{ W  } ^{\frac{1}{3}}\bigg(&{F}^{-(1-\sum p_i)}\,\frac{dr^2}{r^4}+\sum{F}^{+2\,p_i}\,(\sigma^i)^2\bigg),
	\end{aligned}
	\end{eqnarray}
	where the $W(r) $ function (here and hereafter) is given by
	\begin{eqnarray}\label{W}
	\begin{aligned}
	W =\frac{c_3^2-{b}^{2}{F}}{c_3^2-{b}^{2}}{\rm e}^{-\phi_{\infty}}.
	\end{aligned}
	\end{eqnarray}
	If the conditions \eqref{condition1} hold, the
	$W (r)$ is positive everywhere, finite at $r_{H}=-c_1$, and blows up at $r=0$.

	\subsection{Bianchi type $II$}\label{solII}
	In this Bianchi type the $Y_i$ terms of \eqref{newii} are given by
	\begin{eqnarray}\label{}
	\begin{aligned}
	-Y_1=Y_2=Y_3=\frac{1}{2}{g_1 ^2}\left(g_2^2 g_3^2 FU^2\right)^{-1}.
	\end{aligned}
	\end{eqnarray}
	The solutions of \eqref{newii} give the string frame metric \eqref{metric} components as
	\begin{eqnarray}
	g_1 ^2={\rm e}^{-\phi}V^{-1}F^{\frac{1}{2}(p_1-\frac{1}{2})},\label{IIg1}
	\\
	g_2 ^2=l_2^2{\rm e}^{-2\phi}g_1 ^{-2}F^{2p_2},\label{IIg2}
	\\
	g_3 ^2=l_3^2{\rm e}^{-2\phi}g_1 ^{-2}F^{2p_3},\label{IIg3}
	\end{eqnarray}
	where 
	\begin{eqnarray}\label{VII}
	\begin{aligned}
	V=\frac{2 p_1 c_1 q_1}{q_1^2-F^{2p_1}},
	\end{aligned}
	\end{eqnarray}
	and  $p_1$, $p_2$, $q_1$, $l_2$ and $l_3$ are real constants.
	Substituting these solutions into initial value equation \eqref{ini} gives the following constraint on integrating constants
	\begin{eqnarray}\label{iniII}
	\begin{aligned}
	-p_1^2-2\,\left(p_2+p_3\right)-4\,p_2p_3-\frac{1}{4}=0.
	\end{aligned}
	\end{eqnarray}
	To preserve the signature of metric at spatial infinity and near horizon,  the $q_1$ and $p_1$ are required to have opposite signs with $q_1^2-1>0$. Also, using \eqref{con}, the Einstein frame metric is given by
	\begin{eqnarray}\label{metricII}
	\begin{aligned}
	ds^2=-{ W  }^{-\frac{2}{3}}{ F} \,dt^2+{ W  } ^{\frac{1}{3}}V^{-1}F^{-\frac{1}{2}-{p_1}}\bigg(l_2^2l_3^2{F}^{2(p_2+p_3)}\,\frac{dr^2}{r^4}+&V^{2}F^{2p_1}(\sigma^1)^2\\
	&+l_2^2F^{1+2 p_2}(\sigma^2)^2+l_3^2F^{1+2 p_3}(\sigma^3)^2\bigg).
	\end{aligned}
	\end{eqnarray}

	\subsection{Bianchi type $III$}\label{solIII}
	Belonging to    class $B$, this Bianchi type  gives the following constraint equation  by $(r,x^3)$ component of \eqref{classBc}  
	\begin{eqnarray}\label{}
	\begin{aligned}
	\ln \left(g_3 g_1 ^{-1}\right)'=0,
	\end{aligned}
	\end{eqnarray}
	which essentially requires $SO(2)$ isometry with $g_1 =g_3 $.
	Also, the Bianchi type dependent terms in this model are given by
	\begin{eqnarray}\label{}
	\begin{aligned}
	Y_1=Y_3=\left(g_3^2 FU^2\right)^{-1},~~~Y_2=0.
	\end{aligned}
	\end{eqnarray}
	Then, the solutions of \eqref{newii} are
	\begin{eqnarray}\label{IIIg1}
	g_1 ^2=g_3 ^2={\rm e}^{-2\phi}g_2 ^{-2}V^{2}F^{2p_1-1},
	\end{eqnarray}
	\begin{eqnarray}\label{IIIg2}
	g_2 ^2=l_2^2{\rm e}^{-\phi}F^{2p_2},
	\end{eqnarray}
	in which  we have defined
	\begin{eqnarray}\label{Viii}
	\begin{aligned}
	V=\frac{-2 p_1 c_1 q_1}{q_1^2+F^{2p_1}},
	\end{aligned}
	\end{eqnarray}
	and  $p_1$, $p_2$, $q_1$ and $l_1$ are real constants.
	Substituting these solutions into the initial value equation \eqref{ini} gives
	\begin{eqnarray}\label{iniIII}
	\begin{aligned}
	-4\left(p_1^2-p_2^2\right)+2p_2+1=0.
	\end{aligned}
	\end{eqnarray}
	Respecting the signature of metric at infinity and near horizon, the $q_1$ and $p_1$ constants are required to have the same signs. The Einstein frame metric is then given by
	\begin{eqnarray}\label{metriciii}
	\begin{aligned}
	ds^2=&-{ W  }^{-\frac{2}{3}}{ F} \,dt^2+{ W  } ^{\frac{1}{3}}V^2F^{2p_2}\bigg(
	l_2^2V^2{F}^{4(p_1-p_2)-3}\,\frac{dr^2}{r^4}+V^{-2}l_2^2(\sigma^2)^2+F^{-1+2p_1-4p_2}\left((\sigma^1)^2+(\sigma^3)^2\right)\bigg).
	\end{aligned}
	\end{eqnarray}

	
	\subsection{Bianchi type $IV$}\label{solIV}
	In this Bianchi type of class $B$, the constraint equation \eqref{classBc} leads to 
	\begin{eqnarray}\label{}
	{g_3 ^2}{g_1 ^{-2}}=0,\quad \ln\left({g_1 ^2}{(g_2 g_3)^{-1} }\right)'=0,
	\end{eqnarray}
	and so 
	this Bianchi type does not provide any non-singular solution.

	\subsection{Bianchi type $V$}\label{solV}
	
	The Bianchi type dependent terms  are given here by
	\begin{eqnarray}\label{}
	\begin{aligned}
	Y_1=Y_2=Y_3=2\left(g_1^2 FU^2\right)^{-1}.
	\end{aligned}
	\end{eqnarray}
	Also, the following equation is given by the $(r, x^1)$ component of  \eqref{classBc} in this Bianchi type of class $B$
	\begin{eqnarray}\label{}
	\begin{aligned}
	\ln\left({g_1 ^2}{(g_2 g_3)^{-1} }\right)'=0.
	\end{aligned}
	\end{eqnarray}
	It imposes the constraint of  $g_1 ^2=g_2 \,g_3 $, which is satisfied by the following solutions of \eqref{newii}
	\begin{eqnarray}
	g_1^2=VF^{p_1-\frac{1}{2}}{\rm e}^{-\phi},\label{Vg1}
	\\
	g_2^2=F^{2p_2}\,g_1^{2},\quad\quad\label{Vg2}
	\\
	g_3^2=F^{-2p_2}\,g_1^{2},\quad\,\,\,\label{Vg3}
	\end{eqnarray}
	where we have defined
	\begin{eqnarray}\label{VofV}
	\begin{aligned}
	V=\frac{{{{{ c_1}}}{q_1}}p_1}{{F}^{2\,p_1}+{{q_1}}^{2}}.
	\end{aligned}
	\end{eqnarray}
	Here the $p_1$, $p_2$, and $q_1$ are real constants and the $q_1$ and $p_1$ need to have opposite signs.
	The initial value equation \eqref{ini} gives  
	\begin{eqnarray}\label{iniV}
	\begin{aligned}
	-3\,p_1^2+4\,p_2^2-\frac{3}{4}=0.
	\end{aligned}
	\end{eqnarray}
	Also, the Einstein frame metric is given by
	\begin{eqnarray}\label{metricV}
	\begin{aligned}
	ds^2=-&{ W  }^{-\frac{2}{3}}{ F}\,dt^2+{ W  } ^{\frac{1}{3}}V{F}^{-\frac{1}{2}+p_1}\bigg(V^2{F}^{2p_1-2}\,\frac{dr^2}{r^4}+(\sigma^1)^2+{F}^{2p_2}(\sigma^2)^2+{F}^{{-2p_2}}(\sigma^3)^2\bigg).
	\end{aligned}
	\end{eqnarray}

	
	\subsection{Bianchi type $VI_{h}$}\label{solVIh}
	In this Bianchi type we have
	\begin{eqnarray}\label{}
	\begin{aligned}
	(h^2+1)^{-1}Y_1&=(h(h+1))^{-1}Y_2\\
	&=(h+1)^{-1}Y_1=\left(g_1^2 F\,U^2\right)^{-1}.
	\end{aligned}
	\end{eqnarray}
	Also,  
	the $(r, x^1)$ component of  \eqref{classBc} is 
	\begin{eqnarray}\label{}
	\begin{aligned}
	-(1+h)\ln g_1 '+h\ln g_2 '+\ln g_3 '=0,
	\end{aligned}
	\end{eqnarray}
	which imposes the following restriction
	\begin{eqnarray}\label{}
	\begin{aligned}
	\frac{g_3 g_2 ^h}{g_1 ^{h+1}}=\frac{l_2^{h-1}}{l_1^{h+1}},
	\end{aligned}
	\end{eqnarray}
	where $l_1$ and $l_2$ are real constants. In this Bianchi type, the $h=0,1$ cases give rise to  Bianchi types $III$ and $V$, respectively. Excluding these two types, only the $VI_{-1}$  admits  closed  spatial section \cite{closedBianchi}.\footnote{The closed  spatially homogeneous hypersurface  is  compact without boundary.} Since the black hole is usually assumed to have compact horizon \cite{HERVIK20081253}, here we investigate the solution of  the $h=-1$ case which gives the following solutions
	\begin{eqnarray}
	g_1^2=F^{2\,p_1}{\rm e}^{-16\,q_1^{-2}p_2^2F^{-2\,p_2}}{\rm e}^{-\phi},\label{VIhg1}
	\\
	g_2^2=4\,l_2p_2^2c_1q_1^{-1}F^{-p_2-\frac{1}{2}}{\rm e}^{-\phi},\label{VIhg2}
	\\
	g_3^2=l_2^{-2}\,g_2^{2},\quad\quad\quad\quad\quad\label{VIhg3}
	\end{eqnarray}
	where $p_1$, $p_2$, and $q_1$ are real constants. Assuming the $l_2$ to be positive, $q_1$ should be negative. Substituting these solution into the initial value equation \eqref{ini} gives
	\begin{eqnarray}\label{iniVIh}
	\begin{aligned}
	-p_2^2+p_2(4p_1+1)+\frac{3}{4}=0.
	\end{aligned}
	\end{eqnarray}
	The Einstein frame metric is then given by
	\begin{eqnarray}\label{metricVIh}
	\begin{aligned}
	ds^2=-{ W  }^{-\frac{2}{3}}{ F}\,dt^2&+{ W  } ^{\frac{1}{3}}F^{-\frac{1}{2}}\bigg( 16\,p_2^4c_1^2q_1^{-2}{F}^{-\frac{3}{2}+2\,{ p_1}-2\,{ p_2}}{\rm e}^{-\frac{16p_2^2 }{q_1^2F^{2p_2}}}
	\,\frac{dr^2}{r^4}\\
	&+{{4\,c_1}{l_2p_2^2q_1^{-1}}{F}^{-{ p_2}}}\left((\sigma^2)^2+l_2^{-2} (\sigma^3)^2\right)+{F}^{\frac{1}{2}+2\,{ p_1}}{\rm e}^{-\frac{16p_2^2 }{q_1^2F^{2p_2}}}
	(\sigma^1)^2\bigg).
	\end{aligned}
	\end{eqnarray}
	
	\subsection{Bianchi type $VII_{h}$}\label{solVIIa}
	
	In this case, the Bianchi type dependent terms in \eqref{newii} are
	\begin{eqnarray}
	\begin{aligned}
	Y_1&=-\frac{g_1^4-g_2^4}{2\,F\,(g_1g_2g_3U)^2},\\
	Y_2&=\frac{g_1^4-g_2^4}{2\,F\,(g_1g_2g_3U)^2}-\frac{h^2}{F\,g_3^2\,U^2},\\
	Y_2&=\frac{\left(g_1^2-g_2^2\right)^2}{2\,F\,(g_1g_2g_3U)^2}-\frac{h^2}{F\,g_3^2\,U^2}.
	\end{aligned}
	\end{eqnarray}
	Also, the constraint equations given by \eqref{classBc}  are 
	\begin{eqnarray}\label{conviih}
	h\,\ln\left( {g_3}{g_2^{-1}}\right)'=0,\quad h\,(g_2^2g_3^{-2})=0,
	\end{eqnarray}
	which restrict the solutions to be of type  $h=0$.
	In this case, the equations of \eqref{newii}   can be integrated only with $g_1=g_2$ which, leading  to $Y_i=0$, reduces the equations to those of Bianchi type $I$  and consequently  the  solutions can be recovered from there.
	\subsection{Bianchi type $VIII$}\label{solVIII}
	Here the $Y_i$ terms are given by
	\begin{eqnarray}\label{}
	\begin{aligned}
	Y_1=-\frac{1}{2}\left(g_1^2-\left(g_2^2+g_3^2\right)^2\right)\left((g_1g_2g_3U)^2F\right)^{-1},\\
	Y_2=-\frac{1}{2}\left(g_2^2-\left(g_1^2+g_3^2\right)^2\right)\left((g_1g_2g_3U)^2F\right)^{-1},\\
	Y_3=-\frac{1}{2}\left(g_3^2-\left(g_1^2-g_2^2\right)^2\right)\left((g_1g_2g_3U)^2F\right)^{-1}.
	\end{aligned}
	\end{eqnarray}
	Then, the solutions of \eqref{newii} give the metric components of string frame \eqref{metric} as follows
	\begin{eqnarray}
	g_1^2=g_2^2={\rm e}^{-2\,\phi}V_1 ^2F^{2\,p_1-1}g_3^{-2},\label{viig1}
	\\
	g_3^2={\rm e}^{-\phi}V_2 F^{p_2-\frac{1}{2}},\quad\quad\quad\label{viig2}
	\end{eqnarray}
	where
	\begin{eqnarray}\label{Vviii}
	\begin{aligned}
	V_1 =\frac{2\,c_1q_1p_1}{q_1^2+F^{2p_1}},
	\quad
	V_2 =\frac{-2\,c_1q_2p_2}{q_2^2-F^{2p_2}}.
	\end{aligned}
	\end{eqnarray}
	The $p_i$, $q_1$ and $q_2$  are integrating constants, where  $q_2$ and $p_2$ should have the same sign and $q_2^2-1>0$. Also, as a consequence of  initial value equation \eqref{ini}, $p_i$ are subject to the following constraint
	\begin{eqnarray}\label{iniviii}
	\begin{aligned}
	p_2^2-4p_1^2+\frac{3}{4}=0.
	\end{aligned}
	\end{eqnarray}
	Performing the conformal transformation \eqref{con} on \eqref{metric} we get the Einstein frame metric in the following form
	\begin{eqnarray}\label{metricViii}
	\begin{aligned}
	ds^2=&-{ W  }^{-\frac{2}{3}}{ F}\,dt^2+{ W  } ^{\frac{1}{3}} F^{-\frac{1}{2}-p_2} \bigg(V_1 ^4V_2 ^{-1}{F}^{4p_1-2}\,\frac{dr^2}{r^4}+V_2 F^{2p_2} (\sigma^3)^2
	+V_1 ^2V_2 ^{-1}{F}^{2p_1}\big( (\sigma^1)^2+ (\sigma^2)^2\big)\bigg).
	\end{aligned}
	\end{eqnarray}

	\subsection{Bianchi type $IX$}\label{solIX}
	Here, the Bianchi type dependent terms  $Y_i$ are given by
	\begin{eqnarray}\label{}
	\begin{aligned}
	Y_i=\frac{(g_j^2-g_k^2)^2-g_i ^2}{2F(g_1g_2g_3U)^2 }.
	\end{aligned}
	\end{eqnarray}
	where $(ijk)$ is taken to be cyclically as $(123)$. The equations \eqref{newii} can be  integrated  by setting $g_1 =g_3 $. Then, we get
	\begin{eqnarray}
	g_1 ^2=g_3 ^2=V_1 ^2F^{2\,p_1-1}{\rm e}^{-2\,\phi}g_2 ^{-2},\label{ixg1}
	\\
	g_2 ^2=V_2 F^{p_2-\frac{1}{2}}{\rm e}^{-\phi},\quad\quad\quad\label{ixg2}
	\end{eqnarray}
	where
	\begin{eqnarray}\label{viix}
	\begin{aligned}
	V_i =\frac{2c_1q_ip_i}{q_i^2-F^{2p_i}},~i=1,2,
	\end{aligned}
	\end{eqnarray}
	where $q_i$ and $p_i$ are constants and the $q_2$ and $p_2$ are especially required to have opposite signs with $q_2^2-1>0$.
	Substituting these solutions into the initial value equation \eqref{ini} yields
	\begin{eqnarray}\label{iniix}
	\begin{aligned}
	-4\,p_1^2+p_2^2+\frac{3}{4}=0.
	\end{aligned}
	\end{eqnarray}
	Furthermore, the Einstein frame metric reads
	\begin{eqnarray}\label{metricIx}
	\begin{aligned}
	ds^2=&-{ W  }^{-\frac{2}{3}}{ F}\,dt^2+{ W  } ^{\frac{1}{3}}{F}^{-\frac{1}{2}-p_2}\bigg(V_1 ^4V_2 ^{-1}{F}^{4p_1-2}\,\frac{dr^2}{r^4}+F^{2p_2}V_2  (\sigma^2)^2+V_1 ^2V_2 ^{-1}F^{2p_1}( (\sigma^1)^2+ (\sigma^3)^2)\bigg).
	\end{aligned}
	\end{eqnarray}

	\section{Thermodynamic properties of the topological black hole solutions}\label{sec4}
	
	Having found the solutions of low energy string effective action  equations of motion in the previous section,  we come to investigate the physical properties of the solutions. 
	The black hole interpretation of the solutions  with a horizon located at $r_{H}=-c_1$ requires the $p_i$  constants to have some appropriate values such that the  $\tilde{g}_{rr}$  change  its sign crossing the $r_{H}$. Assuming that the solutions are black hole solutions, the relevant values of  $p_i$ will be obtained demanding some special properties  of the solutions.
	Also, the singularity properties of these topological black hole solutions and verification of the first law of thermodynamics will be investigated.
	
	All of the obtained solutions for various  Bianchi types  have finite $\tilde{\gamma}_{\infty}$  in \eqref{trace}. Therefore, as we mentioned earlier, with $c_3\neq -\sqrt{2}b$ the solutions are neither asymptotically flat nor asymptotically (A)dS. Hence, the considering of   Bianchi classes with negative three-dimensional curvature as the horizons has no conflict with the area theorems \cite{Galloway2006}, whose asymptotic flatness condition is violated.

	For calculating the  mass of these non-asymptotically flat solutions we use the Brawn-York formalism which defines the quasi-local conserved mass by
	\cite{PhysRevD471407}
	\begin{eqnarray}\label{quasi}
	M= \frac{1}{\kappa_5^2}\int_{^3B} d^3x\, \sqrt{\tilde{\sigma}}\left(K_{ab}-\tilde{	h}_{ab}K\right)n^a\bar{\xi}^b,
	\end{eqnarray}
	in which $\kappa_5$ is the five-dimensional Newton constant, $^3B$ is the three-dimensional boundary, $n^a$ is the time-like unit normal vector to the boundary  $^3B$,  $K_{ab}$ is  the  extrinsic curvature of the $4$-dimensional boundary $\partial{\cal{M}}$ with induced metric $\tilde{	h}_{ab}$, and $K$ is the trace of $K_{ab}$. 
	We assume that the metrics have a normalized asymptotically
	time-like Killing vector $\bar{\xi}^{\mu}={\rm e}^{-\frac{\phi_{\infty}}{3}}\delta^{\mu}_{t}$  such that $\bar{\xi}^2=\bar{\xi}^{\mu}\bar{\xi}^{\nu}g_{\mu\nu}\rightarrow -1$ for $r\rightarrow \infty$. 
	The $\tilde{\sigma}$ is  determinant of the metric of $^3B$ in Einstein frame where, noting the relation between the coordinate and non-coordinate basis \eqref{vielbin}, we have $\tilde{\sigma}=\tilde{\gamma}| e_{\mu}^{a}|^2$  in which $\tilde{\gamma}$ has been defined by \eqref{sigma} and $| e_{\mu}^{a}|$ is determinant of vielbine matrix whose components for each Bianchi type are presented in Appendix  \ref{app1}.  Then, considering the metric \eqref{metric} along with \eqref{f} and \eqref{sigma}, 
	the mass expression \eqref{quasi} recasts the following general form
	\begin{eqnarray}\label{}
	M=\frac{-c_1}{\kappa_5^2}\lim_{r\rightarrow \infty} {F}^{\frac{1-\lambda}{2}}W ^{\frac{-1}{6}}\left(\frac{d}{dF}\sqrt{\tilde{\gamma}}\right)\omega_3,
	\end{eqnarray}
	where the  volume element $\omega_3=\frac{1}{3!}\int \sigma^1\wedge\sigma^2\wedge \sigma^3$ and the $F(r)$ and $W(r)$ functions have been given by \eqref{45} and \eqref{W}, respectively. Using the obtained solutions with considering the given conditions in \eqref{condition1}, the mass per unit volume for Bianchi type solutions reads
	\begin{eqnarray}\label{mass}
	{\cal{{M}}}=-\frac{c_1{\rm e}^{-\frac{1}{3}\phi_{\infty}}}{\kappa_5^2}\left(\frac{b^2}{2(c_3^2-b^2)}+\alpha-\frac{\lambda}{2}\right),
	\end{eqnarray}
	in which, being different for each Bianchi type,  the $\lambda$ and $\alpha$  are given for each type as
	\begin{eqnarray}
	\text{ $I\&VII_0$:}&& \lambda=2\sum p_i,   \,\, \alpha=0,\label{Ilanda}\\
	\text{ $II$:}&&\lambda=\frac{1}{2}+2 (p_2+p_3)-p_1,  \alpha=\frac{p_1}{q_1^2-1},\label{IIlanda}\\
	\text{ $III$:}&& \lambda=2-4p_1+2p_2,\,\, \alpha=\frac{4 p_1}{q_1^2+1},\label{IIIlanda}\\
	\text{ $V$:}&& \lambda=3 p_1-\frac{3}{2},\,\, \alpha=\frac{3p_1}{q_1^2+1},\label{Vlanda}\\
	\text{ $VI_{-1}$:}&& \lambda=-\frac{3}{2}+2 p_1-2 p_2,\,\, \alpha=\frac{-16 p_2^3}{q_1^2},\label{VIlanda}\\
	\text{ $VIII$:}&& \lambda=4 p_1-p_2-\frac{3}{2}, \,\,\alpha=\frac{4 p_1}{q_1^2+1}+\frac{p_2}{q_2^2-1},\quad\,\,\,\,\label{VIIIlanda}\\
	\text{ $IX$:}&& \lambda=4 p_1-p_2-\frac{3}{2},\,\, \alpha=\frac{4 p_1}{q_1^2-1}+\frac{p_2}{q_2^2-1}.\quad\label{IXlanda}
	\end{eqnarray}
	It is worth mentioning that the quasi-local mass \eqref{mass} coincides with the  mass  obtained  by the Abbott-Deser approach \cite{ABBOTT198276} using the normalized killing vector $\bar{\xi}^{\mu}$.

	Also, the area of horizon  is generally given by
	\begin{eqnarray}\label{area}
	\begin{aligned}
	A_H=\sqrt{\tilde{\gamma}_H}\,\omega_3,
	\end{aligned}
	\end{eqnarray}
	where  $\tilde{\gamma}_H$ is determinant of induced metric \eqref{sigma} on the horizon. A common feature of the obtained Bianchi type solutions is that their $\tilde{\gamma}$ is proportional to $F^{\lambda}$ multiplied by a  factor which is finite on the horizon. Therefore, to have non-zero area of horizon the $\lambda=0$ is required and then the last term in \eqref{mass} vanishes.

	Furthermore, the surface gravity  is defined by\footnote{Here, we need to choose the normalized  killing vector      to have the first law of black hole thermodynamics satisfied.}
	\begin{eqnarray}\label{surfacega}
	\begin{aligned}
	\kappa&=\sqrt{-\frac{1}{2}(\tilde{\nabla}_{\mu}\bar{\xi}_{\nu})\left(\tilde{\nabla}^{\mu}\bar{\xi}^{\nu}\right)}|_{r=r_{H}}\\
	&=\frac{-{\rm e}^{\frac{\phi_{\infty}}{3}}}{2}\sqrt{-\tilde{g}^{rr}(r_H)\tilde{g}^{tt}(r_H)}\tilde{g}_{tt}'(r_H).
	\end{aligned}
	\end{eqnarray}
	The finite and non-zero surface gravity demands finite non-vanishing $\tilde{g}^{rr}\tilde{	g}^{tt}$ on the horizon. Noting  \eqref{f} and \eqref{sigma} and the fact that $\tilde{\gamma}$ is proportional to $F^{\lambda}$, the  $\tilde{g}^{rr}\tilde{	g}^{tt}$ is  proportional to $F^{-\lambda}$. {Hence,  the finite surface gravity requires again $\lambda=0$.}  Then,  the following general form for surface gravity can be obtained
	\begin{eqnarray}\label{surfaceg}
	\begin{aligned}
	\kappa=&\frac{-c_1{\rm e}^{-\frac{\phi_{\infty}}{3}}}{2\sqrt{\tilde{\gamma}_H}}.
	\end{aligned}
	\end{eqnarray}

	Also, the Hawking temperature can be derived from the Euclidean regularity methods \cite{Hawking1983}  in our considered metric by\footnote{The same result can be obtained via the normalized temperature definition    \cite{Brill} 
		\begin{eqnarray}\label{}
		T_{H}(r)=\frac{\sqrt{-(\nabla_{\mu}{\xi}_{\nu})\left(\nabla^{\mu}{\xi}^{\nu}\right)}|_{r=r_{H}}}{4\pi\sqrt{-\xi_{\mu}\xi^{\mu}}},
		\end{eqnarray}
		which is independent of the  normalization of the horizon generator.}
	\begin{eqnarray}\label{Tr}
	\begin{aligned}
	T_H(r)=&\frac{\sqrt{(\tilde{g}_{rr}^{-1}) '\tilde{g}_{tt}'}|_{r=r_{H}}}{4\pi \sqrt{\tilde{g}_{tt}}},
	\end{aligned}
	\end{eqnarray}
	which is infinite on the horizon and has the finite  non-vanishing value at infinity, given by
	\begin{eqnarray}\label{T}
	\begin{aligned}
	T_H
	=\frac{-c_1{\rm e}^{-\frac{\phi_{\infty}}{3}}}{2\sqrt{\tilde{\gamma}(r_H)}}=\frac{\kappa}{2\pi}.
	\end{aligned}
	\end{eqnarray}
	
	Furthermore, the black hole entropy can be obtained using the Wald’s formula \cite{wald}
	\begin{eqnarray}\label{waldf}
	\begin{aligned}
	S=-2\pi\int_{H} d^3x \sqrt{\tilde{\sigma}} \frac{\delta \cal{L}}{\delta \tilde{	R}_{\mu\nu\rho\sigma}}\epsilon_{\mu\alpha}\epsilon_{\rho\nu} =-\frac{\pi}{\kappa_5^2}\epsilon_{\mu\nu}\epsilon^{\mu\nu}A_H,\quad\,\,
	\end{aligned}
	\end{eqnarray}
	which is evaluated on the horizon  and $\epsilon_{\mu\nu}$ is the binormal to the horizon whose normalization is usually chosen as $\epsilon_{\mu\nu}\epsilon^{\mu\nu}=-2$.
	
	Presence of the $3$-form field strength tensor \eqref{H} introduces a charge associated to  the antisymmetric $B$-field. Having no  $H_{t\mu\nu}$ component, the Noether electric charge  of $B$-field is zero, as well as its electric type potential, which is actually defined by the difference of the values of $B_{t\alpha}  $ component of tensor gauge field   at infinity
	and at the horizon \cite{Emparan2004}.
	On the other hand, 
	a topological magnetic charge  can be  defined here by \cite{YOUM19991}  
	\begin{eqnarray}\label{p}
	\begin{aligned}
	Q_m=\frac{1}{\sqrt{2}\kappa_5}\int_{^3B} H=\frac{b}{\sqrt{2}\kappa_5}\omega_3,
	\end{aligned}
	\end{eqnarray}
	whose conservation is associated with the Bianchi identity $dH=0$.   In the following, we will mention this charge by its density  defined by ${\cal{Q}}_m=\frac{b}{\sqrt{2}\kappa_5}$.  
	To calculate the conjugate potential of this magnetic charge, similar to the derivation of electric and magnetic potentials in  Einstein-Maxwell-dilaton theory \cite{Lu2013}, we can use the electric-magnetic duality \cite{EMDuality}. The equations of motion in Einstein frame are invariant under the transformation $(H\rightarrow \tilde{    H}=*H{\rm e}^{-\frac{4\phi}{3}},\phi\rightarrow -\phi)$, where the $*$  stands for Hodge dual operation.
	Considering the field strength tensor \eqref{H}, the $\tilde{    H}$ is a two-form corresponding to an electrically charged Maxwell field 
	\begin{eqnarray}\label{duf}
	\tilde{	H}=\frac{b\,{\rm e}^{2\phi}}{12\,r^2}dt\wedge dr,
	\end{eqnarray}
	whose vector potential
	in a gauge where the  scalar
	potential vanishes on the horizon is  
	\begin{eqnarray}
	\tilde{A}=\left(\frac{c_3}{12 \,b\, W(r)}-\Phi_H\right)dt,
	\end{eqnarray}
	where $\Phi_H=\frac{(c_3^2-b^2){\rm e}^{{\phi_{\infty}}}}{12 \,b\,c_3}$.  The electric charge of this dual vector field, defined by $Q_e=\frac{1}{\sqrt{2}\kappa_5}\int_{^3B} {\rm e}^{\frac{4}{3}\phi_{\infty}} *\tilde{	H}$,  equals  $Q_m$ \eqref{p}.\footnote{ Since in $D$ dimensions for a $p$-form $a$ we have $**a=(-1)^{D-1+p(D-p)} a$, the $Q_m$ and $Q_e$ have the same sign.} 
	Also, the electric potential 
	is  given  by the time component of the dualized tensor gauge potential by
	\begin{eqnarray}\label{du}
	\tilde{\Phi}_e=\bar{\xi}.\tilde{A}|_{r=r_{H}}-\bar{\xi}.\tilde{A}|_{r=\infty}=\frac{1}{\sqrt{2}\kappa_5}\frac{-c_1b\,{\rm e}^{-\frac{\phi_{\infty}}{3}}}{(c_3^2-b^2)}.
	\end{eqnarray}
	Hence, our solutions can be alternatively described by Maxwell electric Hodge dual field to the three form $H$ \eqref{H}.  Now, as a matter of fact that the roles of Maxwell electric and $B$-field magnetic charges are exchanged under the extended Hodge dualization,
	the magnetic potential $\Phi_{m}$ in our solution can be interpreted as the electric potential in the dual frame $ \tilde{\Phi}_e$, which can be rewritten in the following form using \eqref{fiinfini} and \eqref{p} 
	\begin{eqnarray}\label{fib}
	\begin{aligned}
	\Phi_{m}=-\frac{c_1{\cal{Q}}_m{\rm e}^{\frac{2\phi_{\infty}}{3}}}{c_3}.
	\end{aligned}
	\end{eqnarray}

	To check the first  law of black hole thermodynamics, the relation between the solution parameters and physical ones are required. Reminding the conditions \eqref{condition1} where the $c_1$ and $c_3$ have been required to be negative,
	the inverse relations for Bianchi type $I$ are given by
	\begin{eqnarray}\label{c1I}
	-c_1={\rm e}^{\frac{\phi_{\infty}}{3}}\left(\frac{{\cal{Q}}_m^{2}{\rm e}^{\frac{4\phi_{\infty}}{3}}-4\,\kappa_5^2{{\cal{M}}^2}}{2\,{\cal{M}}}
	\right),
	\end{eqnarray}
	\begin{eqnarray}\label{qI}
	-c_3=\frac{{\cal{Q}}_m^2{\rm e}^{\frac{2\,\phi_{\infty}}{3}}}{{\cal{M}}},
	\end{eqnarray}
	where for the other Bianchi types we have\footnote{The $\alpha$ (or equivalently the $q_i$) will be considered as non-varying constants which will be fixed by the first law of thermodynamics. In fact, the $\alpha$  are related to the curvature of three-dimensional space part. This is in analogy with the topological black hole solutions where the constant curvature $k$ of horizon appears in the mass \cite{Aminneborg,Smith,Brill,Yazadjiev}.}
	\begin{eqnarray}\label{c1}
	\begin{aligned}
	-c_{1}=&{\rm e}^{\frac{\phi_{\infty}}{3}}\bigg(\frac{\kappa_5^2{\cal{M}}}{\alpha}-{\frac {2\,{\cal{Q}}_m^{2}{\rm e}^{\frac{4\phi_{\infty}}{3}}}{{\cal{M}}+
			\sqrt {8\, \left( \alpha-\frac{1}{2} \right) \alpha\,\kappa_5^{-2}{\cal{Q}}_m^{2}{\rm e}^{\frac{4\phi_{\infty}}{3}}+{{\cal{M}}}
				^{2}}}}
	\bigg),
	\end{aligned}
	\end{eqnarray}
	\begin{eqnarray}\label{q}
	-c_{3}={\frac {\kappa_5^2{\cal{{\cal{M}}}}+\sqrt {8\, \left( \alpha-\frac{1}{2} \right) \alpha\,{\cal{Q}}_m^{2}{\rm e}^{\frac{4\phi_{\infty}}{3}}+{\kappa_5^2{\cal{M}}}^{2}}}{2\,\alpha \,{\rm e}^{\frac{2\phi_{\infty}}{3}}}}.
	\end{eqnarray}

	Before considering the first law of thermodynamics, let us have a closer look at the dilaton field \eqref{fioa}.
	Having a non-trivial configuration, the dilaton introduces a dilaton charge. Based on  the definition of this charge 
	\cite{garfinkle1991charged}
	\begin{eqnarray}\label{dich}
	D=\int d^3\Sigma^{\mu}\nabla_{\mu}\phi,
	\end{eqnarray}
	where $d\Sigma_{\mu}=\frac{\sqrt{\tilde{\gamma}}}{3!}\epsilon_{\mu\alpha_1\alpha_{3} \alpha_{3}}\sigma^{\alpha_1}\wedge\sigma^{\alpha_2}\wedge \sigma^{\alpha_{3}}$ is the dual of $3$-area form 
	and the integral  is evaluated at spatial infinity, the dilaton charge per volume $\omega_3$ is   given  by 
	\begin{eqnarray}
	\label{Dilatoncharg}
	{\cal{D}}=\frac{c_1b^2}{c_3^2-b^2},
	\end{eqnarray}
	which is  negative holding the conditions \eqref{condition1}. It is worth mentioning that, the same ${\cal{D}}$ can be obtained from the asymptotic expansion of the dilaton field \eqref{fiex} 
	in which the ${\cal{C}}$ is given by
	\begin{eqnarray}
	{\cal{C}}=\frac{c_1^2b^4}{2\left(c_3^2-b^2\right)^2}=\frac{{\cal{D}}^2}{2}.
	\end{eqnarray}
	
	In addition, the Einstein frame effective action \eqref{GBaction} is invariant under the following scaling symmetry in five dimensions
	\begin{eqnarray}\label{shift}
	\phi\rightarrow \phi+\phi_0,\quad B_{\mu\nu}\rightarrow {\rm e}^{-\frac{2}{3}\phi_0}B_{\mu\nu},
	\end{eqnarray}
	which is a constant shift in dilaton accompanied by  a scaling in the gauge $B$-field. 
	When a theory exhibits the shift symmetry of dilaton, the  associated Noether current can be considered to identify whether or not a non-trivial configuration for dilaton is allowed  \cite{Sotiriou,Hui,Kartik}. Especially, this kind of current and its conservation have been used to establish the no-hair theorem in \cite{Hui}, where the $J^{\mu}$ which is given in terms of $\phi'$, respecting the symmetries of metric,  has only the $J^{r}$ as non-vanishing component and the finite value of $J_{\mu}J^{\mu}$ on the horizon requires the $J^{r}$ to vanish on horizon. The last step of their  proof utilizes the
	conservation of $J$, which being rewritten in the form of $\partial_{{\mu}}\left(\sqrt{-\tilde{    g}}J^{\mu}\right)=0$  leads to $\rho^2(r) J^{r}=const$ where the $\rho$ function   measures  the area of 
	spheres with constant $r$. Then, since the $\rho$ remains finite and non-zero even on the horizon, therefore the constant in this equation is zero and so the $J^{r}$ has to vanish everywhere implying that dilaton needs to be constant with forbidding the dilaton hair.  Following these steps  in our considered case with a field strength tensor of \eqref{H} and a $\phi(r)$, we can investigate the related  Noether current to \eqref{shift}  which is given by
	\begin{eqnarray}\label{J}
	J^{\mu}=\partial^{\mu}\phi-\frac{1}{2}{\rm e}^{\frac{4}{3}\phi}H^{\mu\nu\rho}B_{\nu\rho},
	\end{eqnarray}
	and is conserved only on-shell, using the equations of motion \eqref{eq1} and \eqref{eq2}. This current 
	\eqref{J} does not generally respect either the gauge invariance of $B$-field, i.e. $B_{\mu\nu}\rightarrow B_{\mu\nu}+\partial_{{\mu}}\Lambda_{\nu}-\partial_{{\nu}}\Lambda_{\mu}
	$, or the symmetries of the metric (isometries). Although, the later one can be established by imposing an extra condition especially when the $T$-dual solutions are of interest.\footnote{In principle, the invariance of $J^{\mu}$ under the symmetries of metric is not actually indicted by the field equations, but it can be desired as an extra condition. In fact, considering the homogeneous spacetimes, there exist  some    Killing vectors, i.e. ${\cal{L}}_{\chi}g=0$, where ${\cal{L}}_{\chi}$
		is the Lie derivative along the generator of the symmetry
		$\chi$.    Here, the field strength tensor $H$ \eqref{H} and dilaton \eqref{fioa} have the same invariance ${\cal{L}}_{\chi}H={\cal{L}}_{\chi}\phi=0$. Also, the invariance of $\sigma$-model requires  ${\cal{L}}_{\chi}B=d\omega$;
		however, to make the $T$-dual transformation valid, one can use the gauge invariance of  $\sigma$-model under $B\rightarrow B+d\omega$,  
		to choose an adopted coordinate system where all of the background fields are independent of isometry coordinate \cite{GIVEON199477}. In this case, with ${\cal{L}}_{\chi}B=0$ the shift symmetry current \eqref{J} respects the symmetries of metric as well.  
	}
	Also, considering \eqref{H} and noting that the dilaton is  a function of  $r$ only, the presence of the second term in \eqref{J} indicates that besides  $J^r$, there are also non-vanishing components $J^i$.  At the horizon, this current has finite norm  $J_{\mu}J^{\mu}$ and vanishing $J^{r}$.
	However, the presence of $J^{i}$ contests the last mentioned step of proof of \cite{Hui}   where the conservation of $J$  does not require $J^{r}$ to vanish everywhere. 
	Then, having a non-trivial configuration for dilaton field is not forbidden in our considered case. This does not, however, imply that the scalar charge is an independent charge carried by the obtained black holes.
	In some hairy black hole solutions, the regularity condition of $\phi$  and $\phi'$ has been employed to relate the dilaton charge to the mass of black hole implying that the scalar hair is of secondary type \cite{Sotiriou}. {In our case, it just fixed the integrating constant $n$ by \eqref{n}.}
	Considering the mass expression \eqref{mass}  along with aforementioned $\lambda=0$  condition,   the dilaton charge \eqref{Dilatoncharg}  coincides with  the first term in mass \eqref{mass} and  we have 
	\begin{eqnarray}\label{fff}
	{\cal{D}}=-2\left(\kappa_5^2{\rm e}^{\frac{\phi_{\infty}}{3}}{\cal{M}}+\alpha\,c_1\right).
	\end{eqnarray}
	Noting that  the Bianchi type  $I$ model has $\alpha=0$  and  the other Bianchi types with non-zero  $\alpha$  have the relation \eqref{c1} for $c_1$, the relation \eqref{fff} actually shows that
	the dilaton charge can be  expressed in terms of ${\cal{M}}$, ${\cal{Q}}_m$ and the parameter $\phi_{\infty}$. Therefore, the dilaton charge is not an independent characteristic of the obtained black hole solutions and corresponds to a secondary hair \cite{COLEMAN1992175}. 
	Also, considering the     \eqref{p},   \eqref{fib}, and \eqref{Dilatoncharg}, the following relation  is relevant between the dilaton charge and the magnetic charge multiplied by its potential
	\begin{eqnarray}\label{chern}
	{\cal{D}}=-{2}{\kappa_5^2}{\rm e}^{\frac{2}{3}\phi_{\infty}}\Phi_{B}{\cal{Q}}_m.
	\end{eqnarray}
	The horizon radius $r_{H}=-c_1$ in Bianchi type $I$ is given by \eqref{c1I}, but in the other types
	it can be also  rewritten in terms of mass and dilaton charge densities 
	\begin{eqnarray}
	r_{H}=-c_{1}=\frac{1}{2\alpha}\left(2\,\kappa_5^2{\rm e}^{\frac{\phi_{\infty}}{3}}{\cal{M}}+{\cal{D}}\right).
	\end{eqnarray}

	Evidently, the solutions have been characterized by  mass,  magnetic charge $Q_m$ associated to $B$-field and the parameter $\phi_{\infty}$ related to the asymptotic value of dilaton. 
	The dependence of mass on the $\phi_{\infty}$ is one of the unusual features of black hole solutions in dilatonic theories.
	For this type of black hole solutions a cosmological scenario  has been provided in \cite{Gibbons}, where considering the $\phi_{\infty}$ as a varying parameter, the first law of black hole thermodynamics has been modified as 
	\begin{eqnarray}\label{1st}
	d{{M}}=\frac{1}{\kappa_5^2}\kappa \,d A_{H}+\Phi_m d Q_m+\left(\frac{\partial\,M}{\partial \phi_{\infty}}\right)_{{{A_H,Q_m}}}\,d\phi_{\infty},~~~
	\end{eqnarray}
	in which the last term is actually proportional to dilaton charge.  As mentioned in introduction, the nature of dilaton charge, which is not a locally conserved and gauge symmetry protected charge, criticizes this modified version of first law \cite{Astefanesei1,Astefanesei2008}. 
	Here, motivated by \cite{ASTEFANESEI201847}, we consider the case  that besides the gravitational surface term which leads to the quasi-local mass \eqref{quasi}, the effective action is supplemented with a dilaton field boundary term.
	In Hamiltonian formalism where the
	variation of dilaton boundary term is given by \cite{Hertog}
	\begin{eqnarray}\label{fibo}
	\begin{aligned}
	\delta Q_{\phi}[\xi]=-\frac{1}{6\kappa_5^2}\int \bar{\xi}^{\perp}\delta\phi \nabla_{{\mu}}\phi d^3\Sigma^{\mu},
	\end{aligned}
	\end{eqnarray}
	where $\bar{\xi}^{\perp}=\bar{\xi}.n$ with   $n$ the unit normal to space-like surface $\Sigma$ and $\bar{\xi}$ is the normalized time-like killing vector. 
	Considering the obtained metrics, the $d^3\Sigma^{r}$  is proportional to $r^2$ and so there is neither divergent term nor contribution of ${\cal{C}}$ in  $\delta Q_{\phi}$.
	In such a way that the only non-vanishing term in \eqref{fibo} at infinity is given by
	\begin{eqnarray}\label{e}
	\begin{aligned}
	\delta Q_{\phi}[\xi]=-\frac{1}{6\kappa_5^2}{{D}}{\rm e}^{-\frac{\phi_{\infty}}{3}}\delta \phi_{\infty}.
	\end{aligned}
	\end{eqnarray}
	{In particular, the integrablity of this charge needs a functional relationship between $D$ and $\phi_{\infty}$.}\footnote{{In fact, it was first observed  in  AdS context in \cite{Hertog} that   the integrability of  energy in Hamiltonian formalism forces  the ${\cal{D}}$ and ${\cal{C}}$   to be functionally related.}}
	In order to obtain a well defined variational principle  we have to add the boundary term
		\begin{eqnarray}
		S_{\phi}^{ct}=\frac{-1}{6\kappa_5^2}\int_{\partial{\cal{M}}} d^{4}x \sqrt{\tilde{	h}}\frac{Y(\phi_{\infty}){{\rm e}}^{\frac{\phi_{\infty}}{3}}}{\sqrt{\tilde{\gamma}_{\infty}}},
		\end{eqnarray} 
		with boundary condition 
		\begin{eqnarray}\label{dc}
		{\cal{D}}={\rm e}^{\frac{\phi_{\infty}}{3}}\frac{d Y(\phi_{\infty})}{d\phi_{\infty}}.
		\end{eqnarray}
		Noting  \eqref{f} and   the asymptotic expansions of the metric and scalar field   \eqref{fiex}-\eqref{ge},
		it can be easily shown that the variation
		of the action which yields the following boundary term
		\begin{eqnarray}
		\begin{aligned}
		\delta (S_{\phi}+S_{\phi}^{ct})=\frac{1}{6\kappa_5^2}\int_{\partial{\cal{M}}} d^{4}x \sqrt{\tilde{h}}\bigg[&\sqrt{\tilde{	g}^{rr}}\phi'\left(1+\frac{1}{r}\frac{d {\cal{D}}}{d\phi_{\infty}}\right)+\frac{{\cal{D}}}{\sqrt{\tilde{\gamma}_{\infty}}}\bigg]\delta \phi_{\infty},
		\end{aligned}
		\end{eqnarray}
		is well defined at  $r\rightarrow \infty$, i.e. $\lim_{r\rightarrow \infty}	\delta (S_{\phi}+S_{\phi}^{ct})=0$.
		Now, considering \eqref{quasi}, the total energy is given by
		\begin{eqnarray}\label{}
		\begin{aligned}
		E_{tot}&=M+ \frac{1}{6\kappa_5^2}\int_{^3B} d^3x\, \sqrt{\tilde{\sigma}}h_{ab}n^a\bar{\xi}^b\frac{Y(\phi_{\infty}){{\rm e}}^{\frac{\phi_{\infty}}{3}}}{\sqrt{\tilde{\gamma}_{\infty}}}
		\\&=M+\frac{1}{6\kappa_5^2}Y(\phi_{\infty})\omega_3.
		\end{aligned}
		\end{eqnarray}
	On the other hand, the mass \eqref{mass} can be expressed in the following form using the \eqref{area}, \eqref{surfaceg}, and \eqref{Dilatoncharg}\footnote{{It is not a Smarr-like formula and is given here to clarify the variation.
	}}
	\begin{eqnarray}\label{}
	M=\frac{1}{\kappa_5^2}\left(-\frac{1}{2}D {\rm e}^{-\frac{\phi_{\infty}}{3}}+\alpha\kappa A_H \right).
	\end{eqnarray}
	Therefore, if the boundary term of dilaton with condition \eqref{dc} is taken into account, the last term in \eqref{1st} can be concealed  and so the variation of the total energy  in the first law of thermodynamics will be
	\begin{eqnarray}\label{1stnew}
	d{{E}}_{tot}=\frac{1}{\kappa_5^2}\kappa d A_{H}+\Phi_m d Q_m,
	\end{eqnarray}
	which, interestingly,  does not contain the dilaton charge through the variation term of $\phi_{\infty}$.

	In the following subsections, we present the explicit forms of thermodynamic quantities $A_H$ \eqref{area} and $\kappa$ \eqref{surfaceg} for each Bianchi type solution to investigate  the first law of black hole thermodynamics by checking the satisfiability of
	\begin{eqnarray} \kappa=\kappa_5^2\left(\frac{\partial\,{{E}}_{tot}}{\partial A_{H}}\right)_{{{Q_m,\phi_{\infty}}}},\quad \Phi_{m}=\left(\frac{\partial\,{{E}}_{tot}}{\partial Q_m}\right)_{{{A_H,\phi_{\infty}}}},\quad
	\end{eqnarray}
	using the \eqref{c1I}-\eqref{q}. 
	As we have seen before, the following extra condition are required on integrating constants of solutions  to ensure the finiteness of non-zero   area of horizon, entropy, and surface gravity  
	\begin{eqnarray}\label{condition}
	\lambda=0,
	\end{eqnarray}
	where $\lambda$ for any Bianchi type as been  given by \eqref{Ilanda}-\eqref{IXlanda}. It will be shown that these constraints along with the initial value conditions, that have been obtained for each Bianchi type via the equation \eqref{ini}, will fix the $p_i$ constants with compatible values with the black hole interpretation. Also, the singularity behavior of  each Bianchi type solution, the classification of horizon geometries based on the correspondence between  Bianchi types and Thurston geometries, and the extremal condition of the solutions  will be presented.

	\subsection{Bianchi types $I$ and $VII_0$}\label{i}

	The  group of Bianchi type $I$ models is isomorphic to the translation group of three-dimensional Euclidean space $E^3$. 
	This model corresponds to  Thurston's geometries of $R^3, E^2\times R$ and $E^3$   
	with isotropy groups of $e, SO(2)$ and $ SO(3)$, respectively \cite{closedBianchi}. In addition, as we have seen in section \ref{solVIIa}, the constraint equation \eqref{classBc} restricts the $VII_h$ model to have  $h=0$  where the solutions  are equivalent to those of Bianchi type $I$.  Hence, thermodynamic properties of these two types can be investigated in parallel.

	Regarding the initial condition \eqref{iniI} and the  physical requirement of   \eqref{condition}, in which the $\lambda$ for these Bianchi types is given by \eqref{Ilanda}, the allowed real values of the integrating constants are
	\begin{eqnarray}p_i=0,
	\end{eqnarray}
	which means that the generally anisotropic solutions \eqref{Isol} reduce to the isotropic one by imposing the physical requirement \eqref{condition}. Then, 
	in the Einstein frame we have
	\begin{eqnarray}\label{I}
	\begin{aligned}
	ds^2=&-{ W  }^{-{\frac{2}{3}}}{ F} \,dt^2+{ W  } ^{\frac{1}{3}}\left(\frac{dr^2}{r^4\,F}+\sum(\sigma^{i})^2\right).
	\end{aligned}
	\end{eqnarray}
	Also, the  area of horizon \eqref{area} and surface gravity \eqref{surfaceg} take the following forms
	\begin{eqnarray}\label{}
	A_H
	=\sqrt{{c_3}{c_1^{-1}}}\omega_3,
	\end{eqnarray}
	\begin{eqnarray}
	\kappa
	=\frac{-c_1}{2}\sqrt{c_1c_3^{-1}}{\rm e}^{-\frac{\phi_{\infty}}{3}}.
	\end{eqnarray}
	It is straightforward to check that the first law of thermodynamics \eqref{1stnew} is satisfied by these thermodynamic quantities. Also, the Kretschmann and Ricci scalars are proportional to the inverse of $W $, such that
	\begin{eqnarray}\label{riciI}
	R_{\mu\nu\rho\sigma}R^{\mu\nu\rho\sigma}\propto \frac{1}{W ^{\frac{14}{3}}}, \quad
	R\propto \frac{1}{W ^{\frac{7}{3}}}.
	\end{eqnarray}
	So the metric \eqref{I} has two irremovable singularities at $r_1=0$ and $r_2={\frac{c_1b^2}{c_3^2-b^2} }$.  Holding the conditions \eqref{condition1}  the  $r_2$ is negative. Since the metric blows up near $r=0$,  without lose of generality we will study the solutions for $r>0$ and $r_2$ can be ignored. Also, it is worth mentioning that for these Bianchi type solutions the asymptotic value of $\gamma$ in \eqref{trace} is $\gamma_{\infty}={\rm e}^{-\phi_{\infty}}$. 
	
	As we will see in the following, all of the obtained Bianchi type solutions have the $r_1$ and $r_2$ singularities and the given discussion can be applied for all types. However,  there may be other singularities in the some Bianchi types which will be discussed in any cases.

	\subsection{Bianchi type $II$}

	There is a correspondence between Bianchi type $II$ and Thurston's   nilgeometry and Heisenberg group 
	whose
	isotropy groups are $SO(2)$ and $ e$, respectively \cite{closedBianchi}.
	The solutions in this Bianchi type have been found in  subsection \ref{solII}.
	The initial condition \eqref{iniII} along with physical requirement of  \eqref{condition}, in which the $\lambda$ is given by \eqref{IIlanda},  yields
	\begin{eqnarray}\label{consII}
	p_1=\frac{1}{2},~~p_2=p_3=0.
	\end{eqnarray}
	With this fixed constants, the metric \eqref{metricII} recasts the following form
	\begin{eqnarray}\label{II}
	\begin{aligned}
	ds^2=-{ W  }^{-{\frac{2}{3}}}{ F} \,dt^2&+{ W  } ^{\frac{1}{3}}V^{-1}\bigg(l_2^2l_3^2\,\frac{dr^2}{r^4F}+V^{2} (\sigma^1)^2+l_2^2 (\sigma^2)^2+l_3^2\, (\sigma^3)^2\bigg),
	\end{aligned}
	\end{eqnarray}
	where preserving the signature of metric at infinity and near horizon requires   $q_1<0$ and $q_1^2-1>0$. The area of horizon \eqref{area} and surface gravity  \eqref{surfaceg} are also given  by
	\begin{eqnarray}\label{areaii}
	A_H
	=|{l_2l_3c_1^{-1}}|\sqrt{c_3q_1}\,\omega_3,
	\end{eqnarray}
	\begin{eqnarray}
	\begin{aligned}
	\kappa
	=\frac{c_1^2}{2|{l_2l_3}|\sqrt{c_3q_1}}{\rm e}^{-\frac{\phi_{\infty}}{3}},\label{suii}
	\end{aligned}
	\end{eqnarray}
	
	From the first law of thermodynamics \eqref{1stnew} point of view, if the $l_1$ and $l_2$ are taken to be non-varying constants,  the first law  is satisfied with $\alpha=-\frac{1}{4}$. In this case, black hole interpretation of the solution is lost because the $r_{H}=-c_1$ parameter becomes negative. But, a consistent example can be obtained  if one sets $l_2=l_3=\left(-{c_1^2}{q_1}^{-1}\right)^{\frac{1}{4}}$ which satisfies the first law by fixing $\alpha=\frac{1}{4}$, i.e. $q_1=-\sqrt{3}$.
	
	For this Bianchi type solution the asymptotic value of $\gamma$ in \eqref{trace} is $\gamma_{\infty}=-\frac{2c_1}{3}{\rm e}^{-\phi_{\infty}}$.  
	Also, the Ricci and Kretschmann scalars   are proportional to  $ W ^{-\frac{7}{3}}V^{-1}$ and   $ W ^{-\frac{17}{3}}V^{-1}$, respectively. Hence, besides the $r_1=0$ singularity which has been mentioned in Bianchi type $I$, there is another initial singularity at  $r_3=\frac{c_1}{q_1^2-1}$, which  is negative and can be ignored.

	\subsection{Bianchi type $III$}
	The Thurston geometries of $H^2\times E^1$ (where $H^2$ is two-dimensional hyperbolic space) and $ \widetilde{	{SL_2R}}$ locally possess   Bianchi type $III$ symmetry
	with  $ SO(2)$ isotropy \cite{closedBianchi}.\footnote{The $ \widetilde{	{SL_2R}}$ can also have the Bianchi type $VIII$ symmetry.}
	The solutions in this type have been obtained in subsection \ref{solIII}. 
	The initial condition \eqref{iniIII} along with the physical restriction \eqref{condition} in which the $\lambda$ for this type is given in \eqref{IIIlanda}, fixes the constants as
	\begin{eqnarray}p_1=\frac{1}{2},~~p_2=0.
	\end{eqnarray}
	Then, the Einstein frame metric \eqref{metriciii} takes the following form
	\begin{eqnarray}\label{III}
	\begin{aligned}
	ds^2=-{ W  }^{-{\frac{2}{3}}}{ F}\,dt^2&+{ W  } ^{\frac{1}{3}}\bigg(\frac{V^4}{l_2^2r^4\,F}\,dr^2+l_2^2 (\sigma^2)^2+l_2^{-2}V^2( (\sigma^1)^2+ (\sigma^3)^2)\bigg),
	\end{aligned}
	\end{eqnarray}
	where the $V$ function has been defined by \eqref{Viii}. 
	Also, the surface gravity \eqref{surfaceg} and area of horizon \eqref{area} read
	\begin{eqnarray}\label{}
	A_H
	=\frac{\sqrt{c_3c_1^3}}{| l_2|q_1^{2}}\,\omega_3,
	\end{eqnarray}
	\begin{eqnarray}
	\kappa
	=\frac{| l_2|q_1^2}{2\sqrt{c_1c_3}}{{{\rm e}^{-\frac{1}{3}\phi_{\infty}} }}.
	\end{eqnarray}
	Similar to the Bianchi type $II$, considering the  $l_2$   as non-varying constants, the satisfaction of  first law with these thermodynamics quantities needs $\alpha=1$. This value  can not be accepted because leads to non-real $c_1$ and $c_3$ in \eqref{c1} and \eqref{q}. But, for example, choosing $l_2=q_1^{-2}(-c_1)^{\frac{3}{2}}$ the fist law can be satisfied consistently with  $\alpha=\frac{1}{4}$, i.e. $q_1=\sqrt{7}$.

	Here the asymptotic value of $\gamma$ in \eqref{trace} is $\gamma_{\infty}=-\frac{20401c_1}{8}{\rm e}^{-\phi_{\infty}}$.
	Also, investigating the behavior of 
	Ricci and Kretschmann scalars, which are proportional to  $ W ^{-\frac{7}{3}}V^{-5}$ and   $ W ^{-\frac{17}{3}}V^{-11}$, respectively,  shows that  besides the $r_1=0$, 
	there is another initial singularity at $r_3=\frac{-c_1}{q_1^2+1}$, which is not naked since  $r_3<r_H$.

	\subsection{Bianchi type $V$}\label{thermoV}
	
	In isotropic case, the Bianchi type $V$ has hyperbolic geometry $H^3$ with $SO(3)$ isotropy  \cite{closedBianchi}.
	But,   an anisotropic expansion  is not allowed for this Bianchi type  if it is required to admit  a closed spatial section   \cite{closedBianchi}.  
	The solutions  in this Bianchi type have been presented in subsection \ref{solV}, which are initially anisotropic. 
	To have  the black hole interpretation in this model, the metric \eqref{metricV} needs to be isotropic. 
	Interestingly, being consistent with this demand, the solution of the set of initial value equation \eqref{iniV} and 
	physical  condition  \eqref{condition}, in which the $\lambda$ is given by \eqref{Vlanda},  restricts the solutions to be isotropic by 
	\begin{eqnarray}\label{consV}
	p_1=\frac{1}{2},~~p_2=0.
	\end{eqnarray}
	Then, the Einstein frame metric \eqref{metricV} recasts the following form
	\begin{eqnarray}\label{Vf}
	\begin{aligned}
	ds^2=-{ W  }^{-{\frac{2}{3}}}{ F} \,dt^2&+{ W  } ^{\frac{1}{3}}V\bigg(\frac{V^2}{r^4\,F}dr^2+ \sum(\sigma^i)^2\bigg),
	\end{aligned}
	\end{eqnarray}
	where the $V$ function is given by  \eqref{VofV}  and the $q_1$ is required to be negative. Also, the \eqref{area} and \eqref{surfaceg} give
	\begin{eqnarray}\label{araev}
	A_H
	=-c_1\sqrt{2c_3q_1^{-3}}\,\omega_3,
	\end{eqnarray}
	\begin{eqnarray}\label{suv}
	\kappa
	=\sqrt{{2q_1^3}{c_3^{-1}}}{{{\rm e}^{-\frac{1}{3}\phi_{\infty}} }}.
	\end{eqnarray}
	It can be checked that with non-dynamical  $q_1$, 
	the first law of thermodynamics \eqref{1stnew} 
	is satisfied only with $\alpha=\frac{3}{4}$, i.e. $q_1=-1$.

	Here, the asymptotic value of $\gamma$ in \eqref{trace} is $\gamma_{\infty}=\frac{-c_1^3}{8}{\rm e}^{-\phi_{\infty}}$. 
	Also, the Ricci and  Kretschmann scalars are proportional to $ W ^{-\frac{7}{3}}V^{-5}$ and $ W ^{-\frac{17}{3}}V^{-11}$, respectively. Then, considering the $W $ and $V$ functions given by \eqref{W} and \eqref{VofV}, there are initial singularities at $r_1=0$ and $r_3=\frac{-c_1}{q_1^2+1}$, which are hidden behind the horizon since  $r_1<r_3<r_H$.

	\subsection{Bianchi type {$VI_{-1}$}}
	As we have mentioned  in section \ref{solVIh}, the $h=0,1$ cases of $VI_{h}$ are equivalent to Bianchi types $III$ and $V$. Besides them,  only the $VI_{-1}$ case  admits a closed spatial section and is equivalent to the Thurston geometry type of solvegeometry with Abelian isotropy    \cite{closedBianchi}. 
	Here, the integrating constants  $p_i$ are subject to the initial value condition \eqref{iniVIh}. 
	Taking into account the restrictive condition \eqref{condition} along with \eqref{VIlanda}, the constants are constrained to be
	\begin{eqnarray}\label{consVIh}
	p_2=-\frac{1}{2},~~p_1=0.
	\end{eqnarray}
	Then, the Einstein frame metric \eqref{metricVIh} recasts the following form
	\begin{eqnarray}\label{mvi}
	\begin{aligned}
	ds^2=-&{ W  }^{-{\frac{2}{3}}}{ F} \,dt^2+{ W  } ^{\frac{1}{3}}\bigg( c_1^2q_1^{-2}{\rm e}^{-\frac{4 F}{q_1^2}}
	\,\frac{dr^2}{r^4\,F}+{\rm e}^{-\frac{4 F}{q_1^2}}
	(\sigma^1)^2+{c_1}{l_2q_1^{-1}}\left( (\sigma^2)^2+l_2^{-2} (\sigma^3)^2\right)\bigg),
	\end{aligned}
	\end{eqnarray} 
	and the thermodynamic quantities are given by
	\begin{eqnarray}\label{Ahvi1}
	A_H
	=|q_1|^{-1}{\sqrt{c_3\,c_1}}{}\,\omega_3,
	\end{eqnarray}
	\begin{eqnarray}\label{kvh1}
	\kappa
	=|q_1|\sqrt{{{c_1}}{c_3}^{-1}}{{{\rm e}^{-\frac{1}{3}\phi_{\infty}} }}.
	\end{eqnarray}
	Now, considering the mass of this model given by \eqref{mass} and \eqref{VIlanda}, it can be checked that with non-dynamical  $l_2$ and $q_1$, 
	the first law of thermodynamics \eqref{1stnew} 
	can be verified by fixing $\alpha=\frac{3}{4}$, i.e. $q_1=- 2$.
	
	The Ricci and Kretschmann scalars  in these solutions have analogs behavior with the
	Bianchi type $I$ and so there is an initial singularity at the origin $r_1=0$. Also, the asymptotic value of $\gamma$ in \eqref{trace} for this type is $\gamma_{\infty}=\frac{c_1^2}{4}{\rm e}^{-(1+\phi_{\infty})}$.

	\subsection{Bianchi type $VIII$}
	The Thurston type geometry in this Bianchi type is 
	$\widetilde{{SL_2R}}$ and the isotropy groups are $e$ and $SO(2)$ \cite{closedBianchi}. 
	The solutions in this type have been presented in subsection \ref{solVIII}.  Considering the initial value equation \eqref{iniviii} and  the condition \eqref{condition}, with using \eqref{VIIIlanda}, the  constants are fixed as follows 
	\begin{eqnarray}p_1=p_2=\frac{1}{2}.
	\end{eqnarray}
	Then,  the Einstein frame
	metric \eqref{metricViii} recasts the following form
	\begin{eqnarray}\label{vim}
	\begin{aligned}
	ds^2=-{ W  }^{{-\frac{2}{3}}}{ F}\,dt^2&+{ W  } ^{\frac{1}{3}} \bigg(V_1 ^4V_2 ^{-1}\,\frac{dr^2}{r^4F}+V_2  (\sigma^3)^2+V_1 ^2V_2 ^{-1}\big( (\sigma^1)^2+ (\sigma^2)^2\big)\bigg),
	\end{aligned}
	\end{eqnarray} 
	in which the $V_1$ and $V_2$ functions are given by \eqref{Vviii}. The $q_1$ can have any sign but $q_2$ needs to be positive in such a way that $q_2^2-1>0$. Also, the thermodynamic quantities are given by
	\begin{eqnarray}\label{}
	A_H
	={-c_1{q_1^{-2}}\sqrt{-c_3q_2}}\,\omega_3,
	\end{eqnarray}
	\begin{eqnarray}
	\kappa
	=\frac{q_1^2}{\sqrt{-c_3q_2}}{{{\rm e}^{-\frac{1}{3}\phi_{\infty}} }}.
	\end{eqnarray}
	By assuming  $q_1$ and $q_2$ as non-varying parameters,    the first law  of thermodynamics \eqref{1stnew}
	can be confirmed if the $\alpha$ parameter in the mass \eqref{mass}  equals to $\frac{3}{4}$, i.e.  $\frac{2}{q_1^2+1}+\frac{1}{2(q_2^2-1)}=\frac{3}{4}$.

	Furthermore, the
	Ricci  and Kretschmann scalars are proportional to $ W ^{-\frac{7}{3}}V_1 ^{-6}V_2 ^{-1}$ and  $ W ^{-\frac{17}{3}}V_1 ^{-14}V_2 ^{-2}$, respectively. Accordingly, besides $r_1=0$, there are two other initial singularities at 
	$r_3=\frac{-c_1}{q_1^2+1}$, and $r_4=\frac{c_1}{q_1^2-1}$. Obviously, $r_1<r_3<r_H$ and  $r_3$ is not a naked singularity. But, the requirement of $q_2^2-1>0$ implies that the $r_4$ is negative and can be ignored.
	The asymptotic value of $\gamma$ in \eqref{trace} for this type is $\gamma_{\infty}=-\frac{c_1^3q_1^4(q_2^2-1)}{q_2(q_1^2+1)}{\rm e}^{-\phi_{\infty}}$.

	\subsection{Bianchi type $IX$}\label{ix}
	The Bianchi type $IX$ has the spherical
	Thurston  geometry $ S^3$ and  a group isomorphic to  three-dimensional rotation group $SO(3)$ 
	\cite{closedBianchi}. 
	Solutions of this type are given in subsection \ref{solIX}. The initial condition \eqref{iniix} along with condition  \eqref{condition}, in which the $\lambda$ is given by \eqref{IXlanda}, fixes the constants as
	\begin{eqnarray}p_1=p_2=\frac{1}{2}.
	\end{eqnarray}
	With these constants the Einstein frame metric  \eqref{metricIx} gets the form of \eqref{vim}, but the $V_1 $ and $V_2$ functions for this Bianchi type  are given by \eqref{viix}. Also,   the $q_2$ needs to be negative with $q_2^2-1>0$ but $q_1$ can have any sign.   The area of horizon \eqref{area} and surface gravity \eqref{surfaceg} are
	\begin{eqnarray}\label{}
	A_H
	=\frac{-c_1\sqrt{c_3q_2}}{q_1^2}\omega_3,\\
	\kappa
	=\frac{q_1^2}{\sqrt{c_3q_2}}{{{\rm e}^{-\frac{1}{3}\phi_{\infty}} }}.
	\end{eqnarray}
	Checking the first law of thermodynamics \eqref{1stnew}
	with assuming the $q_1$ and $q_2$ as non-varying parameters,  shows that  the first law 
	is satisfied if the $\alpha$ parameter   in  mass expression \eqref{mass} for this Bianchi type equals to $\frac{3}{4}$, i.e. $\frac{2}{q_1^2-1}+\frac{1}{2(q_2^2-1)}=\frac{3}{4}$.
	
	This Bianchi type can admit a constant curvature type horizon, i.e. with $R_{ij}=kg_{ij}$ by constant $k$, if one sets $\pm q_1=q_2$.
	The 
	Ricci and Kretschmann scalars are proportional to  $ W ^{-\frac{7}{3}}V_1 ^{-6}V_2 ^{-1}$ and   $ W ^{-\frac{17}{3}}V_1 ^{-14}V_2 ^{-2}$, respectively. Hence, there are three initial singularity at $r_1=0$, 
	$r_3=\frac{c_1}{q_1^2-1}$, and  $r_4=\frac{c_1}{q_2^2-1}$. Since $q_2^2-1>0$ then $r_3<0$ and can be ignored. However, we don't  have  such a condition on $q_1$ but in the case of $\pm q_1=q_2$ the $r_4$ is negative as well. The asymptotic value of $\gamma$ in \eqref{trace} for this type is then $\gamma_{\infty}=-\frac{39\sqrt{39}c_1^3}{1000}{\rm e}^{-\phi_{\infty}}$.

	Possessing positive Ricci scalar with respect to the induced metric on the horizon, Bianchi type $IX$ model is the only one which is allowed by the horizon theorem \cite{Galloway2006} to be asymptotically flat. This limit can be obtained by setting $c_3=-\sqrt{2}b$, 
	as  indicated by \eqref{trace}. The first law of thermodynamics for this solution is verified with $\alpha=0$. In this case, the quasi-local mass equals the factor $\frac{4}{3}$ times the  Kumar energy obtained by the normalized time-like killing vector $\bar{\xi}$.

	\subsection{ Near-horizon, extremal and asymptotic limits of the solutions}\label{sec5}
	To end this section,  let us investigate some behaviors of the solutions in near-horizon, extremal, and large $r$ limits.
	We have seen  in sections \ref{i}-\ref{ix} that the physical requirement  \eqref{condition} together with the initial value conditions on constants, obtained from  \eqref{ini}, fixed the  $p_i$ constants, identifying the final forms of black hole metrics. Also,  satisfaction of the first law of thermodynamics in Bianchi types $II$-$IX$ constrained the  $\alpha$ (or equivalently the $q_i$) constants to take  special values. Eventually, the  obtained Einstein frame metrics \eqref{I}, \eqref{II}, \eqref{III}, \eqref{Vf}, \eqref{mvi}, and \eqref{vim}    can be written in the following general form 
	\begin{eqnarray}\label{metricE}
	\begin{aligned}
	d\tilde{s}^2
	&=-W^{-\frac{2}{3}}F\,dt^{2}+ \frac{W^{\frac{1}{3}}X}{r^4F}\,dr^{2}+\sum_{i=1}^3\tilde{    g}_{i}^2\,(\sigma^{i})^2,
	\end{aligned}
	\end{eqnarray}
	where the $F(r)$ and $W(r)$  are given by \eqref{45} and \eqref{W}  but the $X(r)$ and $\tilde{	g}_i(r)$ functions are different for any Bianchi types and can be read simply from the obtained metrics. 
	To find the near horizon limit of these metrics we define $\xi=r-r_H$ and so as $r\rightarrow r_H$, i.e. in $\xi\rightarrow 0$ limit, \eqref{metricE} reduces to  
	\begin{eqnarray}
	d\tilde{s}^2=-W_H^{-\frac{2}{3}}\,r_H^{-1}\xi dt^{2}+\frac{W_H^{\frac{1}{3}}X_H} {r_H^3}\,  \frac{d\xi^2}{\xi}+\sum_{i=1}^3\tilde{    g}_{iH}^2\,(\sigma^{i})^2,\quad\,\,
	\end{eqnarray} 
	in which the $W_{H}$, $X_H$, and $\tilde{	g}_{iH}$ are finite values taken by the   $W(r), X(r)$, and $\tilde{	g}_i(r)$ functions on
	the horizon. 
	Then, introducing a new coordinate $x$ by 
	\begin{eqnarray}
	\xi=\frac{r_H^3\,x^2}{4X_HW_H^{\frac{1}{3}}},
	\end{eqnarray}
	we get
	\begin{eqnarray}\label{nhm}
	d\tilde{s}^2=-\left(4W_HX_Hr_H^{-2}\right)^{-1}\,x^2dt^{2}+ dx^2+\sum_{i=1}^3\tilde{    g}_{iH}^2\,(\sigma^{i})^2,\quad\,
	\end{eqnarray} 
	Now, a closer look at the presented  solutions in \ref{i}-\ref{ix} sections reveals that the  factor  $\left(4W_HX_Hr_H^{-2}\right)^{-1}$ for each Bianchi type solution  coincides the square of its surface gravity $\kappa$. Therefore, the $(t,x)$ part of spacetime in the above metric is a flat Minkowski space written in the Rindler coordinates.
	
	In addition, the regularity of the black hole solutions can be investigated to obtain the restriction on the amount of magnetic charge that can be carried by these black holes. From a physical point of view, the required condition is  $r_{H}=-c_1>0$.
	For Bianchi types $I$ and $VII_0$, considering  \eqref{c1I}, the black hole interpretation of the solutions will be lost unless 
	\begin{eqnarray}\label{extermal}
	M^2> \frac{1}{4\kappa_5^2}Q_m^2{{\rm e}^{\frac{4}{3}\phi_{\infty}}}.
	\end{eqnarray}
	Also, for  the Bianchi types $II$ and $III$   the first law of thermodynamics  has been satisfied  with $\alpha=\frac{1}{4}$. In this case, the  real and negative  integration constants $c_1$ and $c_3$, given by \eqref{c1} and \eqref{q}, require again the same condition \eqref{extermal}  to hold.
	The Bianchi types $V$, $VI_{-1}$, $VIII$ and $IX$ satisfied the  first law    with $\alpha=\frac{3}{4}$, where  the integration constants $c_1$ \eqref{c1} and $c_3$   \eqref{q} are  real, but negativity of them is again guarantied by the restriction of type \eqref{extermal}.
	In other words, the  condition \eqref{extermal} gives the extremal condition on the magnetic charge of the all obtained Bianchi type  black hole solutions. For all solutions, in the extremal limit as  $Q_m\rightarrow 2\kappa_5{{\rm e}^{-\frac{2}{3}\phi_{\infty}}}M$, we have $c_3\rightarrow b$ and $r_H=-c_1\rightarrow 0$.
	{In fact similar to Einstein-Maxwell-dilaton solutions \cite{GIBBONS1988741}, the solutions have two horizons at $r=-c_1$ and $r=0$. The  $r=-c_1$ is the event horizon and $r=0$ is generally singular. In extremal case these two horizons coincide at $r=0$.}
	
	Near the horizon in  (near) extremal limit the behavior of temperature  and entropy    depends on the type of Bianchi as follows\footnote{It should be noted that in the extremal limit we have $c_1\rightarrow 0$ and $c_3\rightarrow b$, simultaneously. It can be checked using L'Hospital's rule that the  $\phi_{\infty}$ \eqref{fiinfini} is finite in this limit. } 
	\begin{eqnarray}
	\text{  $I$ $\&$ $VII_0$:}&\quad& T_H\rightarrow 0, S\rightarrow \infty,\label{ex1}\\
	\text{  $II$ $\&$ $III$:}&\quad&T_H\rightarrow 0, S\rightarrow \sqrt{\sqrt{2}\kappa_5{\cal{Q}}_m}/4\,\omega_3,\quad\\
	\text{  $VI_{-1}$:}&\quad& T_H\rightarrow 0, S\rightarrow 0, \label{nn}\\
	\text{  $V$:}&\quad& T_H\rightarrow \sqrt{2\,(\sqrt{2}\kappa_5{\cal{Q}}_m)^{-1}}, S\rightarrow 0,\\
	\text{  $VIII$ $\&$ $IX$:}&\quad& T_H\rightarrow q_1^2 \sqrt{2|q_2|\,(\sqrt{2}\kappa_5{\cal{Q}}_m)^{-1}}, S\rightarrow 0.\quad\quad\label{ex2}
	\end{eqnarray}
	For Bianchi types $I$ and $VII_0$ the behavior is similar to that of Rindler space times \cite{Kabat2014}. 
	In this limits, the $II$ and $III$ types have finite entropy with vanishing temperature,  similar to extreme near horizon Reissner-Nordstrom black hole solution. 
	Thermodynamic analysis    
	of black hole solutions with this behavior has been studied  in  \cite{PhysRevD.88.101503,Hajian2014}.   The $VI_{-1}$ type solution has finite-valued $\kappa/A_H$   in this limit, considering \eqref{Ahvi1} and \eqref{kvh1}. This characteristic, along with  \eqref{nn}, reminds the Extremal Vanishing Horizon (EVH) type black holes 
	\cite{SheikhJabbari2011,Ghodsi2014}.   However, noting  \eqref{mvi}, 
	the $\tilde{g}_i$ part of near horizon metric \eqref{nhm} in this type does not have any zero eigenvalues and hence vanishing of entropy is not accompanied by a vanishing one-cycle on the horizon. Therefore,  the $VI_{-1}$  solution cannot be regarded as an EVH black hole.  Also, for  $V$, $VIII$ and $IX$ type solutions the behavior of temperature and entropy is similar to that of Schwarzschild black holes in the zero-mass limit \cite{Easson2003}. 
	
	{It is worth considering the attractor mechanism \cite{Ferrara1996,Goldstein2005} here to check the consistency of the analysis. The dilaton for  metric \eqref{nhm} is
		\begin{eqnarray}
		\phi(x)=\frac{3}{4}\ln\left(\frac{3c_1c_3}{2\left(b\,c_1x+c_3\right)^2}\right),
		\end{eqnarray}
		and no matter what is the value of $\phi_{\infty}$, near the horizon we have
		$${\rm e}^{{\phi_H}}\equiv\lim_{x\rightarrow 0}{\rm e}^{{\phi(x)}}=\left(\frac{3 c_1}{2c_3}\right)^{\frac{3}{4}}, ~~~~~~{{\phi}}'_H\equiv\lim_{x\rightarrow 0}{{\phi}}(x)'=\frac{3 b\, c_1}{2c_3}.$$ 
		In the extremal limit,  ${{\phi}}'_H$ vanishes, which is a fixed point and  ${\rm e}^{{\phi_H}}=\left(\frac{0}{{2}{\cal{Q}}_m}\right)^{\frac{3}{4}}=0$.
		Also, as discussed in Appendix \ref{app2},   using the equation of motion we can consider $V_{eff}=2\kappa_5^2{\cal{Q}}_m^2{\rm e}^{2\phi}$, where the attractor condition  $\partial_{\phi}V_{eff}=0$ \cite{Goldstein2005} leads again to ${\rm e}^{{\phi_H}}=0$.}\footnote{In comparison with the cases of $V_{eff}=P^2{\rm e}^{\alpha_1\phi}+Q^2{\rm e}^{-\alpha_2\phi}$, where $\alpha_1$ and $\alpha_2$ have the same signs and ${\rm e}^{{(\alpha_1+\alpha_2)\phi_0}}=\frac{\alpha_2Q^2}{\alpha_1P^2}$ extremises the effective potential \cite{Astefanesei2008,Goldstein2005}, here we have $Q=0$ and $P$ proportional to $Q_m$ which, as mentioned before, can be interpreted as electric charge in Hodge dual theory. 
	} Also, using the obtained expression for $\tilde{\gamma}_H$ in \eqref{gamah},  noting \eqref{area} and \eqref{waldf},
	it can be easily checked that  in extremal limit, this formula gives the same behavior for entropy as given in \eqref{ex1}-\eqref{ex2}.

	In addition, using \eqref{fiinfini}, \eqref{p} and \eqref{condition},  the mass of black holes \eqref{mass}  can be rewritten as follows
	\begin{eqnarray}
	{\cal{{M}}}=-\frac{{\cal{Q}}_m^2}{c_3}{\rm e}^{\frac{2}{3}\phi_{\infty}}+\frac{\alpha\, r_H}{{\kappa_D^{2}}}{\rm e}^{-\frac{\phi_{\infty}}{3}}.
	\end{eqnarray}
	In the  high energy limit $r_H\rightarrow 0$ \cite{PANAHIYAN2019388}, which can be alternatively  regarded as  extremal limit here, the
	mass is non zero and is govern by 
	the magnetic charge term,  where in the  asymptotic limit, i.e. as  $r_H\rightarrow \infty$, the dominant term   is the second term which is zero for flat horizon cases.

	It is worth mentioning that although the  metric \eqref{metricE} looks singular as $r\rightarrow \infty$,  its asymptotic scalar curvature  given by \eqref{trace} is finite  in this limit, noting that  the  value of $\tilde{\gamma}_{\infty}$ for each Bianchi type solution given in previous subsections is finite and non-zero. Also, the Ricci scalar \eqref{ricci} on the horizon becomes $\tilde{	R}=-\frac{b^2(c_1c_3^{-1})^{\frac{4}{3}}}{6\tilde{\gamma}_H}$ and $\tilde{\gamma}_H$ is finite for all Bianchi type solutions.
	
	{Furthermore, non-zero components of effective energy momentum tensor \eqref{Teff} are given by
		\begin{eqnarray}
		\begin{aligned}
		T_i^{i\mathrm{(eff)}}=-T_r^{r\mathrm{(eff)}}=\frac{1}{12\tilde{\gamma}}\left(3b^2{\rm e}^{\frac{4\phi}{3}}-2c_1^2F\left(\frac{d\phi}{dF}\right)^2{\rm e}^{-\frac{4\phi}{3}}\right),
		\\
		-T_t^{t\mathrm{(eff)}}=\frac{1}{12\tilde{\gamma}}\left(3b^2{\rm e}^{\frac{4\phi}{3}}+2c_1^2F\left(\frac{d\phi}{dF}\right)^2{\rm e}^{-\frac{4\phi}{3}}\right).
		\end{aligned}
		\end{eqnarray}
		In $r\rightarrow \infty$ limit  
		\begin{eqnarray}\label{ffff}
		\begin{aligned}
		&	T_i^{i\mathrm{(eff)}}=-T_r^{r\mathrm{(eff)}}=\frac{b^2}{12c_3^2\tilde{\gamma}_{\infty}}\left(3c_3^2-2b^2\right){\rm e}^{\frac{4\phi_{\infty}}{3}},\\
		&
		-T_t^{t\mathrm{(eff)}}=\frac{b^2}{12c_3^2\tilde{\gamma}_{\infty}}\left(3c_3^2+2b^2\right){\rm e}^{\frac{4\phi_{\infty}}{3}},
		\end{aligned}
		\end{eqnarray}
		indicating that the pressures and energy density do not diverge at $r\rightarrow \infty$ limit. 
		Also, noting  \eqref{ffff}, or equivalently the \eqref{r},  the asymptotic behavior of solutions can not be regarded as asymptotically (A)dS behavior.  Then, as we have mentioned before, assuming $c_3\neq -\sqrt{2} b$ in \eqref{trace}, we call the solutions  non-asymptotically flat, non-(A)dS topological black hole solutions. 
		On the other hand,  the $\tilde{\gamma}$ of any Bianchi type solutions diverges at its  irremovable singularities, the pressures and energy density blow up at these points.}

	\section{$T$-dual solutions of Bianchi type $V$ black hole}\label{sect}
	In the presence of the isometries of background fields, where the following conditions are satisfied  
	\begin{eqnarray}\label{dd}
	{\cal{L}}_{\chi}g= {\cal{L}}_{\chi}\phi= {\cal{L}}_{\chi}H=0,
	\end{eqnarray}
	with killing vectors  $\chi$, the Buscher's $T$-duality transformations  are valid \cite{BUSCHER1988466}. 
	In this case, a convenient coordinate system can be adopted where all of the background field are independent of the isometry direction $x$ and $ {\cal{L}}_{\chi}=\partial_{x}$   \cite{GIVEON199477}. 
	Then,  the $T$-dual transformation with respect to the isometry direction  $x$ are given by\footnote{Here,  the  \eqref{dualfi} is given based on  our notation for the $\beta$-function equations of \eqref{betaGR}-\eqref{betafR} in which the dilaton is minus twice  of the dilaton of  Ref. \cite{callan1985strings} and \cite{BUSCHER1988466}. }
	\begin{eqnarray}\label{tdual5}
	\bar{{g}}_{xx}=\frac{1}{{g}_{xx}},\bar{{g}}_{x\mu}=\frac{{B}_{{x\mu}}}{{g}_{xx}},\bar{{B}}_{x\mu}=\frac{{g}_{{x\mu}}}{{g}_{xx}},\label{dual1}\\
	\bar{{B}}_{\mu\nu}={B}_{\mu\nu}+\frac{({g}_{x\mu}{B}_{{\nu x}}-{g}_{x\nu}{B}_{{\mu x}})}{{g}_{xx}},\\
	\bar{{g}}_{\mu\nu}={g}_{\mu\nu}-\frac{({g}_{x\mu}{g}_{{x\nu}}
		-{B}_{{x}\mu}{B}_{x\nu})}{{g}_{xx}},\label{t2}\\
	\bar{{\phi}}={\phi}+ \ln{|{g}_{xx}|},\label{dualfi}
	\end{eqnarray}
	where $\bar{{g}}_{{\mu}{\nu}}$, $\bar{{B}}_{{\mu}{\nu}}$ and $\bar{{\phi}}$ are metric, antisymmetric tensor field and dilaton  of the $T$-dual $\sigma$-model.
	
	Here, as an example, we investigate the $T$-dual version of the black hole solutions of Bianchi type $V$. 
	Considering the  vielbeins of this Bianchi type given in the table \ref{table1}  of Appendix \ref{app1}, there are two isometry directions of $x^2$ and $x^3$.  In this respect, the antisymmetric $B$-field associated to field strength tensor \eqref{H} can be considered as\footnote{The \eqref{Bv} is the general form of $B$-field   which leads to the field strength tensor \eqref{H} with ${\cal{L}}_{\chi}B=0$, without including the constant components of $B$ which have no relevant physical interpretation here. 
	}
	\begin{eqnarray}\label{Bv}
	\begin{aligned}
	B=\frac{1}{4}b\,{\rm e}^{2x^1}dx^{2}\wedge dx^3.
	\end{aligned}
	\end{eqnarray}
	Now,  given the dilaton \eqref{fioa}, $B$-field \eqref{Bv}, and the metric \eqref{metric} whose components are given by \eqref{f}, \eqref{45} and \eqref{Vg1}-\eqref{Vg3}, we perform the $T$-dual transformations of \eqref{dual1}-\eqref{dualfi} for example with respect to $x^2$. Then, 
	the $T$-dual dilaton, its asymptotic value, and dilaton charge are obtained as follows
	\begin{eqnarray}\label{dualf}
	\begin{aligned}
	\bar{\phi}&=\ln (V)+2\,x^1+\frac{1}{2}\left(4\,p_2+2\,p_1-1\right)F, \\
	\bar{\phi}_{\infty}&=2\,x^1+\ln \frac{{{{{ c_1}}}{q_1}}}{2({{q_1}}^{2}+1)}, \\
	\bar{\cal{D}}&=\frac{c_1}{2}\left(4\,p_2+2\,p_1-1\right)-\frac{2p_1c_1}{q_1^2+1},
	\end{aligned}
	\end{eqnarray}
	where the $V(r)$ function is given by \eqref{VofV}. Furthermore, all of the components of $T$-dual $B$-field vanish and we have $\bar{B}=0$.  Investigating the properties of  $T$-dual metric, obtained by   \eqref{dual1} and \eqref{t2}, shows that the determinant of the three-dimensional  metric in Einstein frame, $\tilde{\gamma}$,   has been kept invariant.  Here, the non-zero and finite area of horizon and surface gravity in \eqref{surfacega} impose the conditions of $\lambda=0$ and $13\,p_1+8\,p_2-13/2=0$, which along with initial condition \eqref{iniV} fix the integrating  constants again as \eqref{consV}. With these fixed values, the first terms in dilaton and its  charge in \eqref{dualf} vanish and the  $T$-dual string frame metric takes the following form
	\begin{eqnarray}\label{dumetricV}
	\begin{aligned}
	ds^2=&-{ F} \,dt^2+\frac{W V^3}{F\,r^4} {dr^2}+WV(dx^1)^2+\left(WV\right)^{-1}\big({\rm e}^{-2x^1}(dx^2)^2+b\,dx^2dx^3
	+{\rm e}^{2x^1}\left((WV)^2+b^2\right)(dx^3)^2
	\big)
	\end{aligned}
	\end{eqnarray}
	where  $W(r)$ function is 
	\begin{eqnarray}\label{Wd}
	\begin{aligned}
	W =\frac{c_3^2-{b}^{2}{F}}{c_1c_3}.
	\end{aligned}
	\end{eqnarray}
	The Einstein frame metric can be easily obtained by performing the conformal transformation \eqref{con} using $T$-dual dilaton \eqref{dualf}. It can be checked that the $T$-dual solutions in Einstein frame are again hairy black hole solutions where the location of the horizon is unchanged, given by $r_{H}=-c_1$. Here, the second term in shift symmetry current \eqref{J} vanishes but the $x^1$ dependence in the dilaton \eqref{dualf} indicates that the non-vanishing components are  $J^{1}$  and $J^{r}$. This current has a finite norm on the horizon  and,  as a consequence of having the $J^1$ component, its conservation has no conflict with the presence of  dilaton charge. Also, the singularity properties of the black hole have not  changed and the Kretschmann and Ricci scalars of the $T$-dual black hole are divergent at the same non-naked singularities of $r_1=0$ and $r_3=\frac{-c_1}{q_1^2+1}$.

	The invariance of $\tilde{\gamma}$ under $T$-duality shows that the area of horizon and entropy remain invariant under $T$-duality.  
	Also, the mass 
	\eqref{mass} (specialized for Bianchi type $V$ with \eqref{Vlanda}),  surface gravity \eqref{suv} and consequently the temperature \eqref{T} will be rescaled under $T$-duality, but  their final forms are similar to the original ones and just
	rewritten in terms of  $\bar{\phi}_{\infty}$ instead of  $\phi_{\infty}$ as
	\begin{eqnarray}\label{massd}
	{\cal{{\bar{M}}}}=-\frac{c_1{\rm e}^{-\frac{1}{3}\bar{\phi}_{\infty}}}{2\kappa_D^2}\left(\frac{b^2}{c_3^2-b^2}+\frac{3}{q_1^2+1}\right),
	\end{eqnarray}
	\begin{eqnarray}\label{suvd}
	\bar{ \kappa}
	=\frac{1}{|l_2l_3|}\sqrt{\frac{2q_1^3}{c_3}}{{{\rm e}^{-\frac{1}{3}\bar{\phi}_{\infty}} }}.
	\end{eqnarray}
	However, except  $c_1$, the  other constants' role has been changed in $T$-dual version of mass.  The $\bar{B}=0$ indicates that no charge associated with $B$-field is carried by the 
	$T$-dual solutions and the $b$  is just a constant  appeared in three-dimensional part of the metric \eqref{dumetricV}. Also, instead of $c_3$, the $q_1$ is related to  dilaton here and the inverse relations for $c_1$ and $q_1$  are given by
	\begin{eqnarray}
	-c_1= - {\frac {4\,{{\rm e}^{\frac{1}{3}\bar{\phi}_{\infty}} } {\cal{{\bar{M}}}}}{\beta}}+{\frac {12\,{{\rm e}^{\frac{5}{3}\bar{\phi}_{\infty}} }}{\sqrt {{ {\cal{{\bar{M}}}}}^{2}+
				\left( -{\beta}^{2}+6\,\beta \right) {{\rm e}^{\frac{4}{3}\bar{\phi}_{\infty}} }}- {\cal{{\bar{M}}}}}}
	,\quad\,\,
	\\
	-q_1=\beta^{-1}\left( - {\cal{{\bar{M}}}}{{\rm e}^{-\frac{2}{3}\bar{\phi}_{\infty}} }+\sqrt {{ {\cal{{\bar{M}}}}}^{2}{{\rm e}^{-\frac{4}{3}\bar{\phi}_{\infty}} }-{\beta}^{2}+6\,\beta}\right)
	,\,\,\,\quad
	\end{eqnarray}
	in which $\beta=\frac{b^2}{c_3^2-b^2}$. Now, considering $\beta$ as a non-varying constant,
	the first law of thermodynamics
	can be verified with 
	$\beta=\frac{1}{4}$.
	Then, the dilaton charge can be expressed in terms of $T$-dual mass by
	\begin{eqnarray}
	{\cal{{\bar{D}}}}=\frac{-2\,{\rm e}^{\frac{5}{3}\bar{\phi}_{\infty}}}{{\cal{{\bar{M}}}}-\sqrt{{\cal{{\bar{M}}}}^2+5{\rm e}^{\frac{4}{3}\bar{\phi}_{\infty}}}},
	\end{eqnarray}
	which indicates that the $T$-dual solution has also the scalar hair of second kind.

	\section{Conclusion}\label{conclussion}
	
	We have constructed five-dimensional topological black hole solutions at leading order of string effective action in presence of dilaton and antisymmetric $B$-field associated with a magnetic charge.  
	The asymptotic behavior  of the solutions at infinity have been affected by the presence of the magnetic charge such that the solutions are neither asymptotically flat nor asymptotically (A)dS. 
	The solutions have been assumed to have  horizons with Bianchi type symmetries where in the context of violated asymptotic flatness assumption of the horizon geometry theorem  \cite{Galloway2006}, in addition to   flat Bianchi type $I$ and positively curved Bianchi type $IX$, the other negatively curved  Bianchi spaces were also considered.  
	These solutions can be regarded as black hole solutions whose three-dimensional horizons  are modeled on   seven types of  Thurston $3$-geometries corresponding to  Bianchi types, where there exist more possibilities for the horizon geometry than
	the  known constant curvature types of spherical, hyperbolic or flat cases, which  are given by   product constant curvature type $ H^2 \times
	R$ and twisted product types of $\widetilde{{SL_2R}}$, nilgeometry, and solvegeometry.

	Possessing non-trivial configuration, the
	dilaton  field  introduced a dilaton charge in the solutions.
	We gave an  argument based on the Noether current of shift symmetry, demonstrating that this current does not cover all of the assumptions of the no-hair theorem proof of \cite{Hui} and so the presence of dilaton charge has no conflict with the conservation of this current.  The solutions have  dilaton hair which is, however,  of
	secondary type in the sense that the dilaton charge is not an independent charge carried by the black hole solutions and can be entirely determined by the magnetic charge and the mass of the black
	holes. Hence,  the no-hair conjecture still holds.
	Also,  the dilaton charge turned to be proportional to the magnetic charge multiplied by the magnetic potential through equation  \eqref{chern}. It is worth mentioning that a quite similar relation has been obtained  in \cite{Kartik} between the scalar charge and the magnetic charge, accompanied by its potential, in the case of the massless field $\phi$ coupled to the electromagnetic field through the second Chern character by using the Noether current of shift symmetry.

	The mass and other thermodynamic quantities depend in a non-trivial way on the asymptotic value of dilaton $\phi_{\infty}$. The problem with the drastic modification of the first law of (static) hairy black hole thermodynamics which contains the variation of $\phi_{\infty}$ has been considered.  
	Following the idea of \cite{ASTEFANESEI201847}, where a boundary term for dilaton field was included in the action in the quasi-local formalism of mass, we considered the case that the  effective action is also supplemented with a boundary term for dilaton. However, only the spherically symmetric solutions have been considered in \cite{ASTEFANESEI201847}. We applied a similar discussion inspired by the proposed boundary term variation of dilaton in
	\cite{Hertog},
	where with vanishing $\phi_{\infty}$,   the ${\cal{D}}$ and ${\cal{C}}$ coefficients  in the asymptotic expansion of the scalar field were  required to be functionally related for integrability of  energy. This relation which is usually imposed   as a boundary condition, can be fixed  uniquely where  the asymptotic AdS symmetry is of interest. Particularly, the ${\cal{C}}=k{\cal{D}}^2$ with a constant $k$, besides preserving AdS invariance,  is compatible with the conformal symmetry of the boundary \cite{Henneaux2007} and makes the contribution of the scalar field in energy vanish \cite{Anabal2015}. When the conformal invariance  is broken, the trace anomaly leads to an extra contribution of dilaton to the total energy with interesting interpretations \cite{Henneaux2007,Anabalon2016}. For instance,   in dyonic black hole solution, this contribution  leads to satisfactions of the first law  without including an extra scalar charge dependent term  \cite{ASTEFANESEI201847}. In our solutions we are dealing with  non-AdS  cases, however the obtained dilaton  solution satisfies ${\cal{C}}= \frac{{\cal{D}}^2}{2}$ if the  regularity of $\phi$ and $\phi'$ on the horizon is required. The $\phi_{\infty}$ is non-zero here and  $\delta Q_{\phi}$  gives only a term with contribution of $\phi_{\infty}$ and ${\cal{D}}$, which following \cite{Hertog}, have been considered to be functionally related.  Similar to \cite{ASTEFANESEI201847}, it turns out that, the scalar field gives a non-vanishing contribution to the total energy which leads to a well defined variational principle in such a way that	
	even if  the $\phi_{\infty}$ is assumed to vary the first law of thermodynamics does not include the scalar charge dependent term.

	Thermodynamics of the solutions and satisfaction  of the first law of thermodynamics have been investigated for all Bianchi type solutions that admit compact event horizons.  
	The solutions are characterized by three parameters: the mass $M$, the $B$-field magnetic charge $Q_m$, and the constant  $\phi_{\infty}$. We expressed the integrating constants of $c_1$, $c_3$, and $b$ in terms of these parameters where the remaining constants have been fixed by requiring some restrictive conditions. 
	In such a way that, the $n$ and $p_i$ have been fixed by certain conditions, including the initial value condition, the regularity of dilaton on the horizon, and the finite non-vanishing surface gravity and area of horizon in the non-extremal black hole cases. Also, it has been observed that the first law holds for Bianchi type $I$ solutions and for the other types its satisfactions requires fixing the integrating constants of $q_i$.

	The solutions have a horizon hiding
	the scalar curvature singularities including one at the origin for all Bianchi types and another one for some of them, where the later singularity actually appears where the $g_i$ components of the metric vanish. At these points, the Ricci and Kretschmann scalars, the energy momentum invariants $T_{\mu}^{\mu} $ and $T_{\mu\nu}T^{\mu\nu}$, energy density, and pressures of the effective matter, which has been considered to include the contributions of dilaton and $B$-field, blow up.

	The Hodge dual of our solutions can be interpreted as  electrically charged black hole solutions of Einstein-Maxwell-dilaton theory.  The dual transformation keeps the Einstein frame metric fixed but changes the sign of dilaton.  This implies that  the properties that depend on the Einstein metric are not influenced by this transformation (for instance, the singularity behavior and
	thermodynamic proprieties).
	But,  the dilaton charge, which was negative for magnetically charge solutions, is positive for electrically charged ones. 
	It is worth mentioning that for electrically charged black hole solutions the action should be supplemented with a surface term
	which is required to make the variation of the Hamiltonian  well defined  \cite{EMDuality,PhysRevD.73.024015}.
	This surface term is zero in the   magnetic cases but  the variational principle for the gauge field should be considered carefully.

	The extremal and near horizon limit of the solutions has been also studied. We have shown that in near horizon limit $2$-dimensional  Rindler space can be recovered in $(t,r)$ part of spacetime. Also, the  extremal condition appeared to be the same for all Bianchi type black hole solutions so the solutions are regular only if $4\kappa_5^2M^2- {{Q_m^2}}{{\rm e}^{\frac{4}{3}\phi_{\infty}}}>0$. Violation of this inequality changes the nature of the curvature singularity where the naked singularities appear for a sufficiently high magnetic charge.

	Furthermore, $T$-dual transformation is allowed by the symmetries of the considered spacetimes.  As an example,   a class of $T$-dual solutions  is obtained in Bianchi type $V$ which has a non-trivial dilaton and vanishing $B$-field.  
	This solution  admits hairy black hole, where the scalar charge, being completely determined by the black hole mass, is secondary and there is no naked singularity. However, the magnetic charge has not been preserved and no $B$-field charge is carried by the $T$-dual black hole.  Examining the properties and thermodynamics of this solution showed that the location of horizon is unchanged and the area of horizon and consequently the entropy remain invariant under $T$-duality.
	In fact, the $T$-duality invariance of entropy has been pointed out first in \cite{Horowitz1},  where the surface gravity and temperature are invariant as well.
	However, unlike the considered solutions in \cite{Horowitz1}, the $\phi_{\infty}$ is non-zero here and the  mass, surface gravity, and temperature which  depend on ${\phi}_{\infty}$ have been rescaled in such a way that the $T$-dual versions have a similar  expressions to the original ones, but they have been rewritten in terms of asymptotic value of $T$-dual dilaton, $\bar{\phi}_{\infty}$.

	It worth mentioning that families of five-dimensional black hole solutions of gravity theories whose three-dimensional horizons are modeled by some of the eight Thurston geometries have been obtained for instance in \cite{Cadeau,PhysRevD.91.084054,PhysRevD.97.024020,HERVIK20081253}. Especially, the non-trivial geometries which are not constant curvature or  product of constant curvature types hold attention. In this category, the solution with nilgeometry and solvegeometry horizons have been studied, but the case of  $\widetilde{{SL_2R}}$ Thurston geometry, having somewhat more complicated field equations, have been left open. We have found black hole solutions in two Bianchi types $III$ and $VIII$ which correspond to $\widetilde{{SL_2R}}$  geometry. The Lifshitz black hole solutions in these two Bianchi types in Einstein gravity with cosmological constant have been obtained in \cite{Liu2012}. However, differently from their solutions which have zero entropy with non-zero temperature, we have seen that the our solutions admit non-zero entropy with some fixed values of integrating constants.

	The magnetic and scalar charges, masses, and the area of  horizons appeared to have a dependence on the volume of three-dimensional homogeneous part of space, $\omega_3$. This aspect of the solutions  is  similar to that of  four-dimensional topological black hole solutions where some of  thermodynamic quantities are proportional to the area of $\omega_2$ \cite{Brill}. The $\omega_2$ can be completely determined by the topology of  horizon in terms of the Euler characteristic of $2$-manifold. But, a difficulty in the investigating of   $3$-manifolds has been the lack of such a good
	topological invariant, like the Euler characteristic of $2$-manifolds.  In fact,  the  Euler characteristic of  closed $3$-manifolds is zero \cite{thurston1983three}. However, in some cases the $w_3$ can be regarded as a topological invariant  and further work is under progress in this sense.

	\section*{Acknowledgments}
	The authors would like to thank M. M. Sheikh-Jabbari for his valuable discussions and comments. 
	This research was supported by Iran National Science Foundation (INSF) under grant No. 96011219.

\appendix
\section{Bianchi  type classification}\label{app1}
In this Appendix,  the classification of real  three-dimensional Bianchi  type Lie algebras are presented. 
The relation between coordinate and non-coordinate basis is given by
\begin{eqnarray}\label{vielbin}
\sigma^i={e_\alpha}^i(x) \,dx^{\alpha},
\end{eqnarray}
where  $\{\sigma^{i},~i=1,2,3\}$  are left invariant basis $1$-forms and  ${e_\alpha}^i(x)$ are  vielbeins.  For each Bianchi type the associated vielbeins,  corresponding Thurston type geometries and the isotropy groups are presented in table $1$, in which  for Bianchi type $VII_h$ we have defined $X={\rm e}^{-kx^1}\cos(a\,x^1)$, $Y=-\frac{1}{a}{\rm e}^{-kx^1}\sin(a\,x^1)$, $k=\frac{h}{2}$, and $a=\sqrt{1-k^2}$.

Considering the metric \eqref{metric},  non-zero components of  Riemann  and Ricci tensors  are given by
\begin{eqnarray}\label{rnon}
\begin{aligned}
R^i_{~jkl}&=\Gamma_{l j}^{m}\Gamma_{km}^{i}-\Gamma_{kj}^{m}\Gamma_{lm}^{i}-f_{kl}^{~~m}\Gamma_{mj}^{i},\\
R^r_{~jrl}&={\Gamma_{lj}^{r}}'+\Gamma_{l j}^{m}\Gamma_{rm}^{r}-\Gamma_{rj}^{m}\Gamma_{lm}^{r},\\
R^r_{~trt}&={\Gamma_{tt}^{r}}'+\Gamma_{tt}^{r}\Gamma_{rr}^{r}-\Gamma_{rt}^{t}\Gamma_{tt}^{r},\\
R^r_{~jkl}&=\Gamma_{l j}^{m}\Gamma_{km}^{r}-\Gamma_{kj}^{m}\Gamma_{lm}^{r}-f_{kl}^{~~m}\Gamma_{mj}^{r},
\end{aligned}
\end{eqnarray}
\begin{eqnarray}\label{ricnon}
\begin{aligned}
R_{ij}&={\Gamma_{j i}^{r}}'
+\Gamma_{j i}^{e}\Gamma_{d e}^{d}+\Gamma_{j i}^{r}\Gamma_{d r}^{d}+\Gamma_{j i}^{r}\Gamma_{r r}^{r}
-\Gamma_{d i}^{e}\Gamma_{j e}^{d}-\Gamma_{d i}^{r}\Gamma_{j r}^{d}-\Gamma_{r i}^{e}\Gamma_{j e}^{r}
-f_{d j}^{~~e}\Gamma_{e i}^{d},\\
R_{rj}&=\Gamma_{j r}^{e}\Gamma_{d e}^{d}-\Gamma_{d r}^{e}\Gamma_{j e}^{d},\\
R_{rr}&={\Gamma_{r r}^{r}}'-\Gamma_{jr}^{i}\Gamma_{r r}^{j}-\Gamma_{t r}^{t}\Gamma_{rt}^{t},
\end{aligned}
\end{eqnarray}
where the prime symbol stands for derivative with respect to $r$ and $f_{ij}^{~k}$ are the structure constants of Bianchi Lie algebras. Also, the connection coefficients are given by
\begin{eqnarray}\label{gnon}
\begin{aligned}
&\Gamma_{ij}^{k}=-g^{d k}(f_{i d}^{~~e}g_{e j}+f_{j d}^{~~e}g_{e i})+f_{ij}^{~~k},\\
&\Gamma_{ir}^{i}=\ln g_{ii} ',~~\Gamma_{tr}^{t}=\Gamma_{rt}^{t}=\frac{1}{2}\ln F '\\
&\Gamma_{ij}^{r}=-\frac{1}{2}FU^2{g_{ij}}',~~\Gamma_{rr}^{r}= \ln\left( FU^2 \right) ',~~\Gamma_{tt}^{r}=\frac{1}{2}F\,U^2F'.
\end{aligned}
\end{eqnarray}
The Riemann  and Ricci tensors in coordinate basis can be obtained by  multiplying the vielbeins $e_{\alpha}^{~i}$, for example $R_{\alpha\beta}=e_{\alpha}^{~i}e_{\beta}^{~j}R_{ij}$ and $R_{\alpha0}=e_{\alpha}^{~i}R_{i0}$.

\leftmargin=0cm
\begin{center}
	{\bf Table 1: Bianchi-type  classification}
\end{center}
\begin{tabular}{l l l l  l l l p{9mm} }\hline
	Type&Class&$d\sigma^i=\frac{1}{2}f^i_{jk}\sigma^j\sigma^k $&
	Coordinate basis&Geometry&Isotropy \\
	\hline \hline
	
	& & $d\sigma^1=0$& &$E^3$& $SO(3)$\\
	I&A&$d\sigma^2=0$&$ \sigma^i=dx^i$& $E^2\times R$&$SO(2)$\\
	& & $d\sigma^3=0$& & $R^3$&$e$ \\ \hline
	& & $d\sigma^1=\sigma^2\wedge\sigma^3$&$\sigma^1=dx^2-x^1dx^3$& \\
	II&A&$d\sigma^2=0$&$\sigma^2=dx^3$&Heisenberg  G&$e$\\
	& & $d\sigma^3=0$&$ \sigma^3=dx^1$&Nilgeometry&$SO(2)$ \\ \hline
	& & $d\sigma^1=0$&$\sigma^1=dx^1$&  \\
	III&B&$d\sigma^2=0$&$\sigma^2=dx^2$&$H^2$$\times E^1$&$SO(2)$\\
	& & $d\sigma^3=\sigma^1\wedge\sigma^3$&$\sigma^3={\rm e}^{x^1}dx^3$&  $\widetilde{{SL_2R}}$&$SO(2)$ \\ \hline
	& & $d\sigma^1=\sigma^1\wedge\sigma^3+ $&&&& \\&&
	$\sigma^2\wedge\sigma^3$&
	$\sigma^1={\rm e}^{-x^1}dx^2-x^1{\rm e}^{-x^1}dx^3$&& &\\
	IV&B&$d\sigma^2=\sigma^2\wedge\sigma^3$&$\sigma^2={\rm e}^{-x^1}dx^3$
	&---&\\
	& & $d\sigma^3=0$&$\sigma^3=dx^1$& &\\ \hline
	& & $d\sigma^1=0$&$\sigma^1=dx^1$& & &\\
	V&B&$d\sigma^2=\sigma^1\wedge\sigma^2$&$\sigma^2={\rm e}^{x^1}dx^2$
	&$H^3$&$SO(3)$\\
	& & $d\sigma^3=\sigma^1\wedge\sigma^3$&$\sigma^3={\rm e}^{x^1}dx^3$&  \\  \hline

	\multirow{3}{*}{	$VI_{a}$ } &  & $d\sigma^1=0$& $\sigma^1=dx^1$& \\
	& B & $d\sigma^2=h\sigma^1\wedge\sigma^2$& $\sigma^2=
	{\rm e}^{hx^1}dx^2$ & &  \\
	&  & $d\sigma^3=\sigma^1\wedge\sigma^3$&$\sigma^3={\rm e}^{x^1}dx^3$ \\
	$VI_{-1}$& & & &Solvgeometry &  $e$& \\
	\hline
	
	\multirow{3}{*}{	$VII_{h}$ 
	} &  & $d\sigma^1=-\sigma^2\wedge \sigma^3$& $\sigma^1=(X-k\,Y)dx^2-Ydx^3$& \\ 
	& B & $d\sigma^2=\sigma^1\wedge\sigma^3$& $\sigma^2=
	Y\,dx^2+(X+kY)dx^3$ &&  \\
	&  &~~+$h\sigma^2\wedge\sigma^3$, $d\sigma^3=0$&$\sigma^3=dx^1$ \\
	$VII_0$ & A & & & $E^3$&  $SO(3)$& \\
	\hline
	
	\multirow{3}{*}{VIII} &  & $d\sigma^1=\sigma^2\wedge\sigma^3$& $\sigma^1=dx^1
	+((x^1)^2-1)dx^2$&&&&\\
	&&&$+(x^1+x^2-x^2(x^1)^2)dx^3$&  $\widetilde{{SL_2R}}$&$e$ \\
	& A & $d\sigma^2=-\sigma^3\wedge\sigma^1$& $\sigma^2=dx^1
	+((x^1)^2+1)dx^2$&&&&\\
	&&&$+(x^1-x^2-x^2(x^1)^2)dx^3$ &   $\widetilde{{SL_2R}}$& $SO(2)$ \\
	&  & $d\sigma^3=\sigma^1\wedge\sigma^2$&$\sigma^3=2x^1dx^2+(1-2x^1x^2)dx^3$   \\ \hline
	\multirow{3}{*}{IX} &  & $d\sigma^1=\sigma^2\wedge\sigma^3$& $\sigma^1=-\sin x^3dx^1+\sin x^1\cos x^3dx^2$&$SU(2)\approx S^3$&$e$  \\
	& A & $d\sigma^2=\sigma^3\wedge\sigma^1$& $\sigma^2=\cos x^3dx^1+\sin x^1 \sin x^3dx^2$ &  $SU(2)\approx S^3$&$SO(2)$  \\
	&  & $d\sigma^3=\sigma^1\wedge\sigma^2$&$\sigma^3=\cos x^1dx^2+dx^3$& $S^3$&$SO(3)$   \\ \hline

\end{tabular}

\section{Effective potential}\label{app2}
Here, we rewrite the equations of motion \eqref{22}-\eqref{eq2} to derive the expression of the effective potential, $V_{eff}$ in attractor mechanism. We have
\begin{eqnarray}
\tilde{R}_{\mu\nu}	=\frac{1}{3}\tilde{\nabla}_{\mu}\phi\tilde{\nabla}_{\nu}\phi+\frac{{\rm e}^{\frac{4\phi}{3}}}{4}(H_{\mu\kappa\lambda}H_{\nu}^{\kappa\lambda}-\frac{2}{9}H^2\tilde{g}_{\mu\nu}),\label{rrr}
\\
\frac{1}{\sqrt{\tilde{	g}}}\partial_{{\mu}}\left(\sqrt{\tilde{	g}}\partial^{{\mu}}\phi\right)=\frac{{\rm e}^{\frac{4\phi}{3}}}{12}H^2,\label{6}\\
{\partial}_{\mu}\left(\sqrt{\tilde{	g}}{\rm e}^{\frac{4\phi}{3}}H^{\mu}_{~\nu\rho}\right)=0.\label{}
\end{eqnarray}
Using \eqref{rrr} we have
\begin{eqnarray}\label{}
\tilde{	R}_{tt}=\frac{\tilde{    g}_{tt}}{3\tilde{\gamma}}V_{eff}(\phi),\quad
\tilde{	R}_{ii}=\frac{\tilde{    g}_{ii}}{6\tilde{\gamma}}V_{eff}(\phi),
\end{eqnarray}
where 
\begin{eqnarray}\label{}
V_{eff}(\phi)={2\kappa_5^2}{}{\rm e}^{\frac{4\phi}{3}}Q_m^2.
\end{eqnarray}
By definition \cite{Goldstein2005}, the $V_{eff}$ can be regarded as  “effective potential ” for the scalar field, where from \eqref{6} we get
\begin{eqnarray}\label{}
r^2\partial_{{r}}(r^2F(r)\partial_r\phi)=\frac{1}{4}\partial_{\phi}V_{eff}(\phi).
\end{eqnarray}
Also, the $rr$ component of \eqref{rrr}  gives
\begin{eqnarray}\label{}
\begin{aligned}
2r^4&\tilde{    g}_{tt}{\rm e}^{-\frac{2\phi}{3}}(\frac{1}{2}\sum_{i<j} (\ln \tilde{	g}_i^2)'(\ln \tilde{	g}_j^2)'-\frac{1}{3}\phi')+r^4(\tilde{    g}_{tt})'\ln(\tilde{\gamma})'{\rm e}^{-\frac{2\phi}{3}}+V_{eff}(\phi)+\tilde{	Y}=0.
\end{aligned}
\end{eqnarray}
On the horizon $\tilde{    g}_{tt}$ vanishes and  $r^2\tilde{    g}_{tt}'=-c_1{\rm e}^{\frac{2\phi_H}{3}}$. Consequently
\begin{eqnarray}\label{gamah}
\tilde{\gamma}_{H}=\frac{-c_1^2\frac{d\tilde{\gamma}}{dF}|_{r=r_{H}}}{V_{eff}(\phi_H)+\tilde{	Y}|_{r=r_{H}}},
\end{eqnarray}
in which for different Bianchi types we have
\begin{eqnarray}
\text{ $I\&VII_0$:}&& \tilde{	Y}=0,\\
\text{ $II$:}&&\tilde{	Y}=V^2, \\
\text{ $III$:}&& \tilde{	Y}=2l_2^2V^2,\\
\text{ $V$:}&&\tilde{	Y}= 12 \,V^2,\\
\text{ $VI_{-1}$:}&& \tilde{	Y}=c_1^2,\\
\text{ $VIII$:}&& \tilde{	Y}=4V_1^2+V_2^2,\\
\text{ $IX$:}&&\tilde{	Y}=4V_1^2+V_2^2. \label{}
\end{eqnarray}
where with given metrics  in subsections \ref{i}-\ref{ix}, the $V$ function for each Bianchi types can be found in \eqref{VII}, \eqref{Viii}, \eqref{VofV}, \eqref{Vviii}, and \eqref{viix}.


\bibliographystyle{h-physrev}
\bibliography{blackhole22}
\end{document}